\documentclass[aps,prl,reprint,superscriptaddress]{revtex4-2} 

\usepackage{graphicx}
\usepackage{dcolumn}
\usepackage{bm}

\begin{document}

\preprint{APS/123-QED}

\title{Nanoscale magnetism probed in a matter-wave interferometer}

\author{Yaakov Y. Fein}
\email{yaakov.fein@univie.ac.at}
\affiliation{University of Vienna, Faculty of Physics, Vienna Center for Quantum Science and Technology (VCQ), Boltzmanngasse 5, A-1090 Vienna, Austria\\}
\author{Sebastian Pedalino}
\affiliation{University of Vienna, Faculty of Physics, Vienna Center for Quantum Science and Technology (VCQ), Boltzmanngasse 5, A-1090 Vienna, Austria\\}
\affiliation{University of Vienna, Vienna Doctoral School in Physics, Boltzmanngasse 5, A-1090 Vienna, Austria\\}
\author{Armin Shayeghi}
\affiliation{University of Vienna, Faculty of Physics, Vienna Center for Quantum Science and Technology (VCQ), Boltzmanngasse 5, A-1090 Vienna, Austria\\}
\author{Filip Kia\l{}ka}
\affiliation{University of Vienna, Faculty of Physics, Vienna Center for Quantum Science and Technology (VCQ), Boltzmanngasse 5, A-1090 Vienna, Austria\\}
\author{Stefan Gerlich}
\affiliation{University of Vienna, Faculty of Physics, Vienna Center for Quantum Science and Technology (VCQ), Boltzmanngasse 5, A-1090 Vienna, Austria\\}
\author{Markus Arndt}
\affiliation{University of Vienna, Faculty of Physics, Vienna Center for Quantum Science and Technology (VCQ), Boltzmanngasse 5, A-1090 Vienna, Austria\\}

\date{\today}

\begin{abstract}
We explore a wide range of fundamental magnetic phenomena by measuring the dephasing of matter-wave interference fringes upon application of a variable magnetic gradient. The versatility of our interferometric Stern-Gerlach technique enables us to study alkali atoms, organic radicals and fullerenes in the same device, with magnetic moments ranging from a Bohr magneton to less than a nuclear magneton. We find evidence for magnetization of a supersonic beam of organic radicals and, most notably, observe a strong magnetic response of a thermal C$_{60}$ beam consistent with high-temperature atom-like deflection of rotational magnetic moments. 
\end{abstract}

\maketitle
\paragraph{Introduction.---\hspace{-3mm}} 
Magnetism, from the quantized deflection of atoms in the Stern-Gerlach experiment~\cite{gerlach22} to bulk ferromagnetism~\cite{vanVleck32}, is quantum mechanical at heart. This relationship is symbiotic: magnetic phase shifts in neutron interferometry were essential in demonstrating a number of fundamental quantum phenomena~\cite{allman93,wagh97,rauch75,werner75}, and magnetic gradients have been employed as coherent beam splitters in matter-wave interferometry~\cite{machluf13}. 

Here, we use matter-wave interferometry to measure magnetic properties of the interfering particles themselves. We apply tunable magnetic gradients within the Long-baseline Universal Matter-wave Interferometer (LUMI)~\cite{fein19,fein20c} and monitor the response of the interference fringes. Unlike in classical beam deflectometry, where one measures the deflection and/or broadening of a macroscopic beam profile, the presence of nanoscale interference fringes allows us to resolve nanometer-level deflections and forces as small as $10^{-26}$ N~\cite{fein20}. While measuring the envelope phase shift of the fringes is suitable for induced dipole moments~\cite{gerlich21}, monitoring the fringe visibility enables us to study species with permanent dipole moments with relaxed phase stability requirements.

In a three-grating Talbot-Lau interferometer like LUMI~\cite{hornberger12}, a near-field interference pattern is imprinted into the molecular beam density behind the second grating. This takes the form of a near-sinusoidal modulation with periodicity $d$ if the gratings are separated by near-multiples of the Talbot length, $L_T = d^2/\lambda_{\text{dB}}$, with $\lambda_{\text{dB}}$ the de Broglie wavelength and $d$ the grating periodicity. Transversely scanning the third grating while monitoring the transmitted flux reveals the interference fringes. Talbot-Lau interferometry is a robust technique with good mass scalability and lenient coherence requirements~\cite{clauser97}, making it particularly attractive for the measurement of molecular properties. 

The universality of the interferometry scheme enables us to study a variety of species with vastly different magnetic properties, from alkali atoms to organic molecules. In the case of atoms, visibility modulation upon application of a magnetic gradient is due to the dephasing and rephasing of the atomic hyperfine substates in the unpolarized beam, similar to previous atom interferometry experiments~\cite{schmiedmayer94,Jacquey07}. The magnetic phase accumulation for isolated molecules is more subtle due to additional degrees of freedom such as vibrational and rotational modes. At low vibrational temperatures, spins can be locked to a molecular axis~\cite{deHeer10}, while internal spin relaxation~\cite{khanna91} and avoided crossings in the Zeeman manifold due to spin-rotation coupling~\cite{xu05} can lead to Langevin-like paramagnetism and one-sided deflection in a magnetic gradient. Such effects complicate the interpretation of molecular Stern-Gerlach experiments~\cite{knickelbein04}, but at the same time provide access to the richer physics of molecular magnetism~\cite{coronado19}. 

\begin{figure}
    \includegraphics[width=\columnwidth]{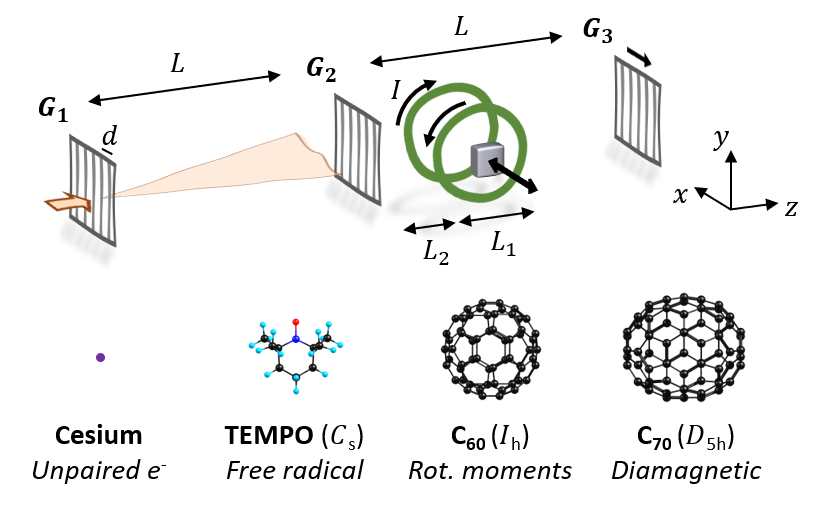}
    \caption{\label{fig:sketch} The interferometer with three equidistantly-spaced gratings $G_{1-3}$ and the magnetic interaction region between $G_2$ and $G_3$. The latter consists of an anti-Helmholtz coil pair for the atomic experiments and a permanent magnet for the molecular experiments. $G_3$ is scanned along the $x$-axis to mask the sinusoidal interference fringes before mass selection and detection of the transmitted flux. The species used in this study are shown with their corresponding point groups and the dominant magnetic behavior observed in our scheme.}
\end{figure}

The key features of our setup are sketched in Fig.~\ref{fig:sketch}. Knudsen cells were used to produce thermal atomic and fullerene beams, while a pulsed valve was used to create a supersonic beam of organic radicals. The interferometer consists of three gratings of period $d=266$~nm equidistantly spaced by $L = 0.98$~m. Three nanomechanical gratings were used for the atomic experiments, while an optical phase grating formed by a retro-reflected 532~nm laser beam was used as the central grating for the molecules~\cite{gerlich08}. 

We apply magnetic gradients across the molecular beam to yield a differential magnetic phase shift for the interferometer paths. For the atomic experiments we employed anti-Helmholtz coils, similar to Ref.~\cite{Jacquey07}. The zero-field region at the coil center was avoided to prevent non-adiabatic spin flips. For the molecular experiments we required stronger magnetic gradients, so we instead used a permanent rare-earth magnet that could be translated transversely to the molecular beam. The lengths $L_{1,2}$ defined in Fig.~\ref{fig:sketch} extend beyond the physical extent of the coils/permanent magnet, taking fringe fields into account (see Supplemental Material~
\footnote{see Supplemental Material, which includes Refs.~\cite{hornberger03,cheeseman96,gauss96,wilson05,kendall92}, for details on the characterization and modeling of the gradient field regions, additional experimental and data analysis details, and computational details for the rotational magnetic moments.}). Characterizations of the magnetic fields as well as the modeled field gradients for both the coils and the permanent magnet are provided in the Supplemental Material. We find good agreement with theoretical models for both systems~\cite{bergeman87,camacho13}. 

\paragraph{Atomic theory.---\hspace{-3mm}}
Atoms with non-zero spin exhibit Zeeman splitting in a magnetic field $B$, and an atomic beam is thus symmetrically deflected in a magnetic gradient~\cite{gerlach22}. Our atomic experiments are conducted in the weak-field regime ($B < 100$~G), and are therefore sensitive to the hyperfine structure rather than merely the electronic spin. We can write the semi-classical force~
\footnote{While the magnetic force itself is essentially classical (i.e., a phase gradient causing an envelope phase shift of the interference pattern) and the semi-classical formalism is used throughout, one obtains identical results for the phase shift with a path integral approach~\cite{Storey94}.}
on a given hyperfine substate $m_F$ as
\begin{equation}
\bm F = -\nabla (-\bm \mu \cdot \bm B) = m_F g_F \mu_{\text{B}} \nabla B,
\end{equation}
with $\bm \mu$ the magnetic moment, $g_F$ the Landé g-factor and $\mu_{\text{B}}$ the Bohr magneton. Here we have implicitly assumed adiabatic following of $\bm \mu$ along $\bm B$. The transverse component of the force yields an envelope phase shift of the interference fringes~\cite{cronin09}. For a given longitudinal velocity $v$ and substate $m_F$, this phase shift is
\begin{equation}
\label{eq:phi}
\phi_{m_F} = \frac{2 \pi}{d} \Delta x_{m_F} = \frac{2 \pi}{d} \frac{m_F g_F \mu_{\text{B}}}{mv^2}(C+C_0),
\end{equation}
with
\begin{equation}
    \label{eq:C}
C = \int_0^{L_1} \int_0^z \frac{\partial B}{\partial x} \,dz \,dz' + L_{\text{drift}}\int_0^{L_1} \frac{\partial B}{\partial x} \,dz.
\end{equation}
Here, ${L_{\text{drift}} = L-L_1-L_2}$, and $C_0$ is a static contribution due to a background gradient $\partial B_0/\partial x$ (see Supplemental Material). The first term corresponds to the deflection within the force region, while the second term accounts for the displacement that accumulates over the remaining drift length, assuming a constant forward velocity. For the anti-Helmholtz coils, we find $C=10.3$~G\,m for a current of 1~A. 

The measured interference pattern is the sum of the $N$ individual hyperfine interference patterns averaged over the velocity distribution $\rho(v)$. Considering the symmetry of the $\pm m_F$ deflections, the visibility $V$ can be written as
\begin{equation}
\label{eq:r}
V = \frac{V_0}{N} \left| \int_0^\infty \rho(v) A(v) \sum_{F,m_F} \cos [\phi_{m_F}(v)] \,dv \right|.
\end{equation}
Here, $V_0$ is the maximum visibility in the absence of magnetic gradients and $A(v)$ is a weak dependence of the visibility amplitude on velocity~\cite{nimmrichter08} which we neglect for the atomic and radical experiments. The hyperfine structures of the studied isotopes are provided in the Supplemental Material.

\begin{figure}
    \includegraphics[width=\columnwidth]{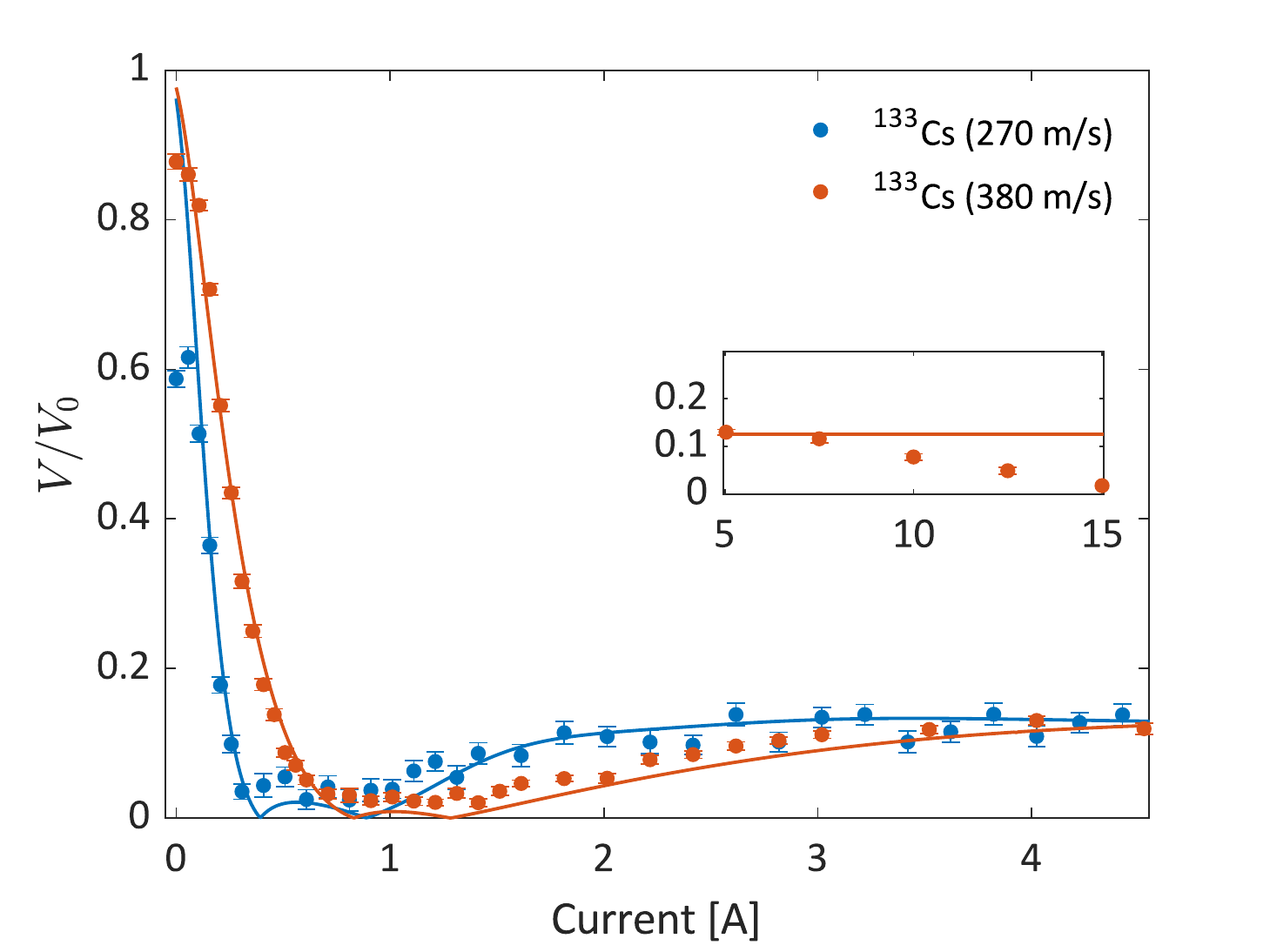}
    \caption{\label{fig:data_cs} Cesium interference visibility as a function of anti-Helmholtz coil current. The current-dependent $C$-factor (Eq.~\ref{eq:C}) used in the solid theory curves is determined by the measured coil geometry. We fit an additional phase contribution due to a background gradient (see main text). The inset shows the high-current behavior, with the line indicating the expected asymptote. Each point consists of multiple interference scans, and error bars are standard errors. Visibilities are normalized using the asymptotic value rather than the zero-current point (see Supplemental Material).}
\end{figure}

\paragraph{Cesium results.---\hspace{-3mm}} 
In Fig.~\ref{fig:data_cs}, we show the interference visibility of $^{133}$Cs as a function of the anti-Helmholtz coil current, for two different velocity distributions. To a good approximation, all magnetic substates $m_F$ are equally populated in the thermal atomic beam. As the magnetic gradient is increased, the visibility decreases as the interference fringes of the various hyperfine states are interspersed, until a value at which each pair of hyperfine substates $\pm m_F$ is mutually deflected by a multiple of $d$, at which point the patterns overlap and the visibility revives. The broad velocity distribution washes out these revivals, leaving an asymptote corresponding to the number of non-magnetic ($m_F=0$) substates divided by the total number of hyperfine substates $N$ (2/16 for $^{133}$Cs). 

There is excellent agreement between the data and Eq.~\ref{eq:r} when we include a small background field gradient of 0.4~G/m, as determined by a least-squares fit. The deviation from theory at currents below 0.15~A is consistent with residual magnetic fields along the flight path, while the drop at currents above 4.5~A is likely non-magnetic in origin. See the Supplemental Material for further details as well as additional rubidium data. 

\paragraph{Molecular theory.---\hspace{-3mm}}
For atoms and molecules with neither nuclear nor electronic spin, strong magnetic gradients can still yield observable deflections~\cite{mairhofer18,fein20,fein20b} due to induced magnetic moments $\bm \mu= m \chi_m \bm B/\mu_0$, with $m$ the particle mass and $\chi_m$ the mass susceptibility. The resulting, typically diamagnetic~\cite{vanVleck32}, deflection of the interference pattern is one-sided, causing visibility loss with increasing $(\bm B \cdot \nabla) B_x$ due to the finite velocity spread~\cite{fein20,fein20c}.

Molecular beam dynamics are richer than for atomic beams, since there are typically a number of excited molecular body rotations with quantized angular momentum $\bm R$ and corresponding magnetic moments $\mu_{\text{rot}}$. For molecules with both spin and rotational angular momentum in an external magnetic field, one must consider spin-rotation coupling in addition to the usual Zeeman interaction terms~\cite{flygare74}. The resulting Zeeman splitting thus depends on whether $B$ is strong enough to decouple spin and rotation or if a coupled basis is more appropriate. Since the rotational $g$-factor ($g_{\text{rot}}$) is typically much smaller than the spin $g$-factor ($g_{\text{spin}}$), the Zeeman splitting is usually dominated by spin in the strong-field regime~\cite{knickelbein04}. Our molecular experiments are in the intermediate to strong-field regime, in contrast to the weak-field atomic experiments with the anti-Helmholtz coils.

Stern-Gerlach experiments on polyatomic species with spin and rotational degrees of freedom can exhibit one-sided rather than symmetric deflection as obtained for atoms. This has been observed in metal cluster beams~\cite{cox85,deHeer90,bucher91}, which exhibit a time-averaged projection of the magnetic moment onto the magnetic field axis of the form $\mu_{\text{eff}} \propto \mu^2 B/k_{\text{B}} T$. This corresponds to the low-field/high-temperature limit of the Langevin function, typically associated with bulk paramagnetism. In molecular beams, the numerical pre-factor of $\mu_{\text{eff}}$ and the relevant temperature depend on whether the spin is locked to the molecular framework or can thermally fluctuate~\cite{payne07,deHeer10}. The origin of magnetization in an isolated molecular system has been the subject of debate and has been explained via both a superparamagnetic model~\cite{khanna91,knickelbein01,payne07} and an avoided crossing model arising from spin-rotation coupling~\cite{xu05,xu08,rohrmann13}.

The response of molecular interference fringes to a magnetic gradient depends on whether the deflection is symmetric or one-sided. For atom-like symmetric deflection, the theory is analogous to Eq.~\ref{eq:r}, while for one-sided deflection one expects monotonic visibility loss due to averaging over the velocity distribution, given by
\begin{equation}
    \label{eq:r_OS}
    V =  V_0 \left| \int_0^\infty \rho(v) A(v) \exp [i\phi(v)] \,dv \right|,
\end{equation}
where $\phi \propto \mu_{\text{eff}}$ for magnetized molecules or $\phi \propto \chi_m B$ for a diamagnetic response. 

\begin{figure}
    \includegraphics[width=\columnwidth]{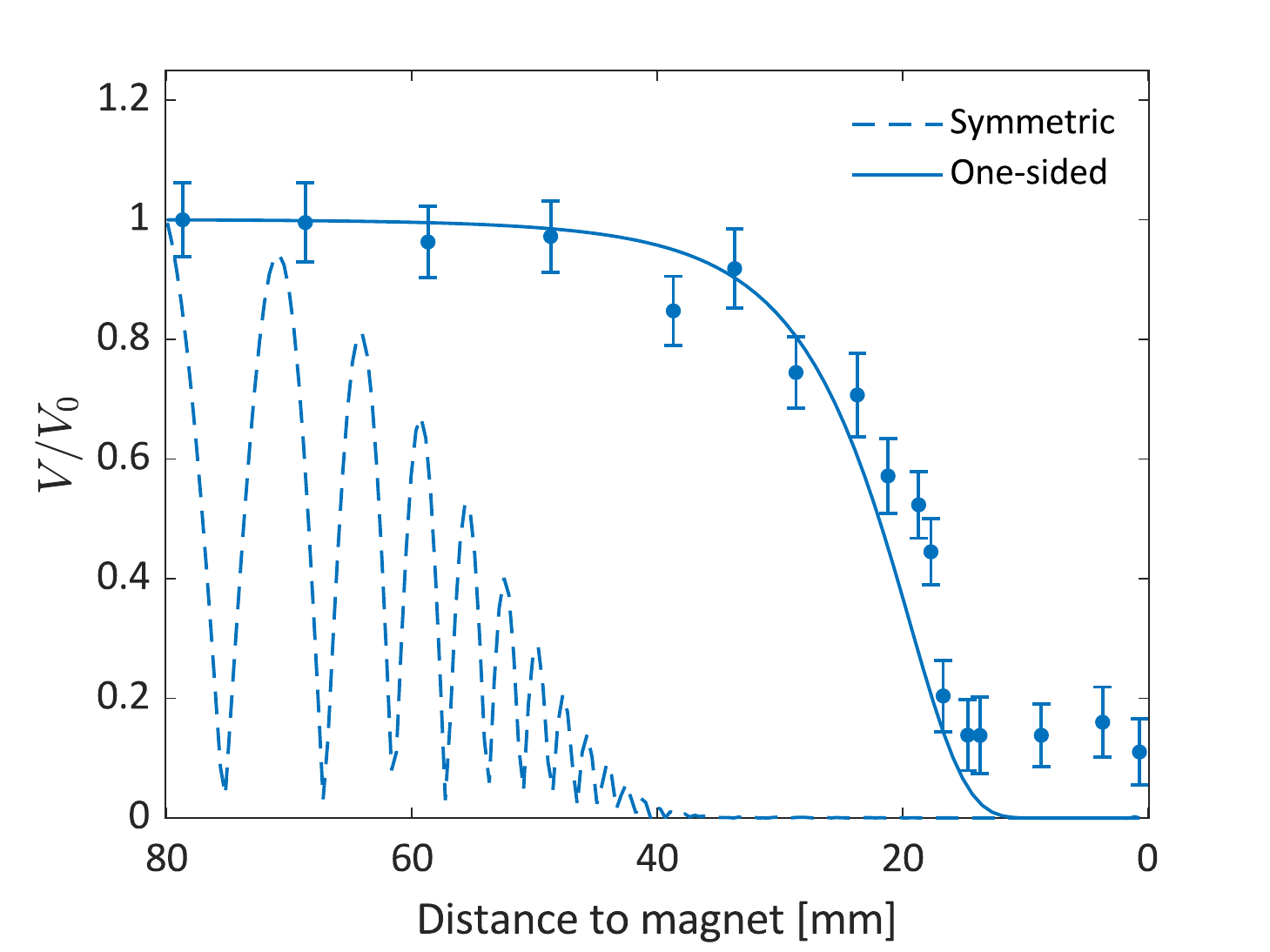}
    \caption{\label{fig:data_tempo} TEMPO interference visibility as a function of magnet distance. A one-sided deflection model assuming ${\mu_{\text{eff}}=0.1\mu_{\text{B}}}$ (solid line) gives better agreement than a symmetric deflection model of the unpaired electronic spin (dashed line). The non-zero baseline of the data is likely a fitting artifact, also seen for the C$_{60}$ isotopomers in the Supplemental Material. Visibilities are normalized to the first data point (magnet withdrawn) and error bars are standard errors.}
\end{figure}

\paragraph{Organic radical results.---\hspace{-3mm}}
The organic radical TEMPO ((2,2,6,6-Tetramethylpiperidin-1-yl)oxyl) is often used as a spin label in electron spin resonance spectroscopy and has also been studied in Stern-Gerlach experiments~\cite{amirav83,gedanken89}. A molecular beam was formed by supersonic expansion from a pulsed valve~\cite{even00} (see Supplemental Material) with a Gaussian velocity distribution centered at 694~m/s and a spread of only 23~m/s. The response of TEMPO interference fringes to the permanent magnet is shown in Fig.~\ref{fig:data_tempo}. Analogous experiments with the TEMPO dimer indicated qualitatively similar behavior, albeit with poorer statistics due to the lower beam flux. 

The dashed line shows a symmetric deflection model in which the response is determined exclusively by the unpaired electron in the strong-field regime (since $g_{\text{spin}} \gg g_{\text{rot}}$, see earlier discussion). We obtain much better agreement with a one-sided deflection model assuming ${\mu_{\text{eff}}=0.1\mu_{\text{B}}}$ in Eq.~\ref{eq:r_OS} (solid line). This is more than predicted by a Langevin-like response, even using the smaller rotational temperature $T_{\text{rot}}$, as would be appropriate for a locked spin~\cite{deHeer10}. The discrepancy could be explained by TEMPO being in an intermediate regime with some degree of Langevin-like magnetization (potentially due to spin-rotation coupling~\cite{xu05,xu08}), but also exhibiting residual symmetric deflection~\cite{rohrmann13}. Such an interpretation is consistent with previous experiments~\cite{gedanken89}, in which indications of asymmetric splitting were tentatively attributed to an avoided crossing model.

\begin{figure}
    \includegraphics[width=\columnwidth]{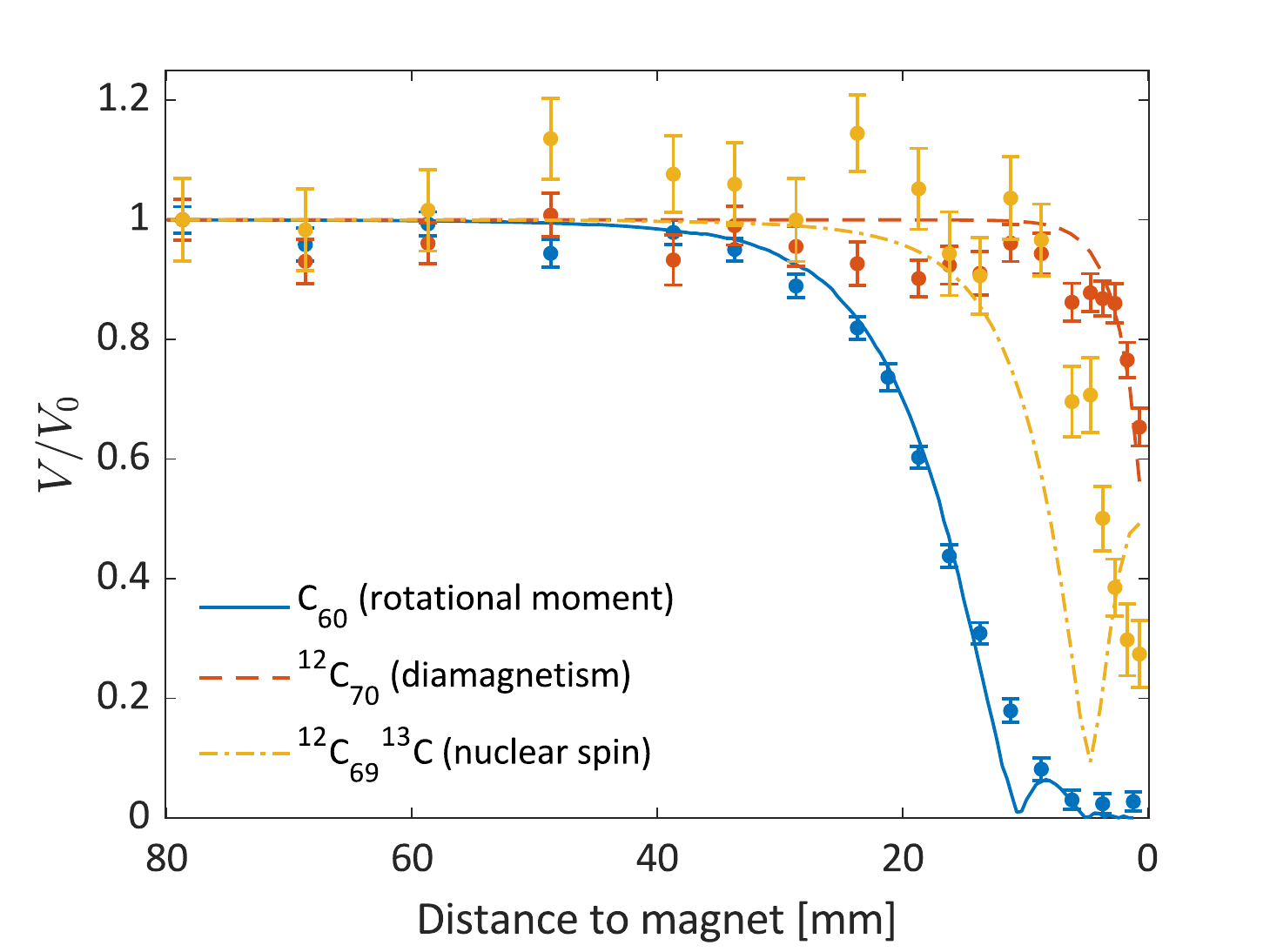}
    \caption{\label{fig:data_fullerenes} C$_{60}$ (all isotopomers), $^{12}$C$_{70}$ and $^{12}$C$_{69}$$^{13}$C interference visibility as a function of magnet distance. There are three distinct magnetic phenomena on display here: one-sided diamagnetic deflection of $^{12}$C$_{70}$, symmetric deflection of the nuclear spin of $^{12}$C$_{69}$$^{13}$C and symmetric deflection of rotational moments of C$_{60}$. Visibilities are normalized to the first data point (magnet withdrawn) and error bars are standard errors. There are no free parameters in the theory curves.}
\end{figure}

\paragraph{Fullerene results.---\hspace{-3mm}}
The response of C$_{60}$ and C$_{70}$ interference visibility to the permanent magnet is shown in Fig.~\ref{fig:data_fullerenes}. The weakly diamagnetic nature of these fullerenes in bulk is well established~\cite{Haddon95}. The $^{12}$C$_{70}$ response agrees well with the predicted visibility loss due to one-sided diamagnetic deflection. The dashed theory curve has no free parameters, using only the literature diamagnetic susceptibility value of C$_{70}$ and the empirical velocity distribution. An observed phase shift of the interference fringes away from the magnet further confirms the expected diamagnetic response. 

With its unpaired nuclear spin, $^{12}$C$_{69}$$^{13}$C exhibits a more rapid loss of fringe visibility than $^{12}$C$_{70}$. The behavior can be described by one-sided diamagnetic deflection in combination with symmetric deflection of the $^{13}$C magnetic moment (dash-dotted line), assuming a strong-field response dominated by the nuclear spin. The moderate disagreement with theory can be partially explained by the drifting $V_0$ level, but could also imply that an intermediate-field treatment is more appropriate. We also note that the stronger response of $^{12}$C$_{69}$$^{13}$C compared to $^{12}$C$_{70}$ implies an absence of Langevin-like magnetization, which would predict complete relaxation of the nuclear spin (since $k_{\text{B}} T \gg \mu^2 B$).

The most surprising finding in this measurement series was the strong magnetic response of C$_{60}$, as seen for all isotopomers (see Supplemental Material). With no electronic spin or orbital angular momentum (or even nuclear, in the case of $^{12}$C$_{60}$), we expected to observe an even weaker diamagnetic response than $^{12}$C$_{70}$. Instead, we observed a visibility drop which would imply a susceptibility two orders of magnitude larger than the literature diamagnetic value. 

Since the visibility is completely washed out at high fields, the underlying mechanism must apply to the majority of C$_{60}$ molecules in the beam, ruling out rare transient events like triplet-excitation in the optical grating. We also ruled out the possibility that deformed cage structures could have led to dangling bonds by measuring a single NMR peak at 142.8 ppm for both un-sublimated and sublimated C$_{60}$ samples~\cite{kim03}. Langevin-like magnetization plays no role here in the absence of spin-rotation coupling~\cite{xu08}, and would in any case be negligible for the hot fullerenes, where ${k_{\text{B}} T \gg \mu^2 B}$.

Instead, we attribute the behavior to the symmetric deflection of rotational magnetic moments. The rotational moment of a spherical top molecule is given by ${\mu_{\text{rot}} = M g_{\text{rot}}\mu_{\text{N}}}$~\cite{eshbach52}, with $M$ the projection of the rotational angular momentum $R$ onto a space-fixed axis. We calculate $g_{\text{rot}}$ with density functional theory at the B3LYP/def2-QZVPP level of theory~\cite{becke93,weigend05} using the Gaussian 16 program~\cite{frisch16} (see Supplemental Material). We find an isotropic value of ${g_{\text{rot}}=-0.0141}$ ($\mu_{\text{N}}$ units), which, for the most-occupied rotational state ${R_{\text{max}}=329}$~
\footnote{The $R$ distribution is proportional to the product of the degeneracy and the Boltzmann factors. The degeneracy of spherical top molecules like C$_{60}$ is broken due to ro-vibrational effects~\cite{bunker99}, giving a degeneracy of $2R+1$ (as for symmetric tops) rather than $(2R+1)^2$. The C$_{60}$ rotational constant is $\tilde B$ = 0.0028~cm$^{-1}$ ($\tilde A$ = 0.0022~cm$^{-1}$ and $\tilde B$ = 0.0019~cm$^{-1}$ for C$_{70}$).},
gives a maximum magnetic moment projection of 4.6~$\mu_{\text{N}}$.

C$_{60}$ thus behaves as an atom in a high spin state, with the substitution ${m_Fg_F\mu_{\text{B}} \rightarrow Mg_{\text{rot}}\mu_{\text{N}}}$. The visibility response can be calculated using Eq.~\ref{eq:r}, replacing the summation over $F, m_F$ with an integration over $M$ from $-R_{\text{max}}$ to $R_{\text{max}}$~
\footnote{Averaging over the $R$ distribution at 870~K gives comparable results to using $R_{\text{max}}$, which is computationally more efficient.}.
This gives the solid theory curve in Fig.~\ref{fig:data_fullerenes}, which agrees well with the measured response.

To understand why C$_{70}$ shows no comparably strong effect of rotational moments ($^{12}$C$_{70}$ is consistent with a diamagnetic response alone), one must consider the difference in molecular symmetry. For a prolate symmetric top like C$_{70}$, there is the additional projection of $R$ onto the molecular symmetry axis, given by the quantum number $K$. The ensemble average of the different $K$ projections, combined with the small $g_{\text{rot}}$ values (see Supplemental Material) reduces the magnitude of the effect. Moreover, in the absence of field-induced alignment (valid here, since $k_{\text{B}} T_{\text{rot}} \gg \mu_{\text{rot}} B$), the projection of $\bm R$ onto $\bm B$ varies due to the changing direction of $\bm B$ along the molecular beam (the projection $M$ is constant only along a space-fixed axis, see Supplemental Material). This reduces the magnetic phase accumulated by symmetric top molecules with rotational magnetic moments in our scheme. Such an effect is not relevant for a spherical top like C$_{60}$, where, in the absence of a unique molecular axis, $\bm R$ adiabatically follows $\bm B$ even as it rotates in the space-fixed frame, in complete analogy with spin in atoms.

Rotational moments similarly do not play a strong role for an asymmetric top like TEMPO, the magnetic response of which is dominated by its unpaired electronic spin. Additionally, $T_{\text{rot}}$ of order 10~K are typical in such supersonic expansions~\cite{amirav80}, yielding much smaller $R_{\text{max}}$ than the thermal fullerene beams and hence much smaller $\mu_{\text{rot}}$. Its $g_{\text{rot}}$ values lie between those of the fullerenes (see Supplemental Material).

\paragraph{Conclusion.---\hspace{-3mm}}
Our interferometric Stern-Gerlach technique offers significantly improved spatial and force resolution compared to classical beam deflectometry and enables us to measure magnetic effects spanning orders of magnitude in strength in the same device. We have observed the symmetric splitting of alkali atoms, the magnetization of an organic radical beam and the weak one-sided diamagnetic deflection of C$_{70}$. Most intriguingly, we observed a strong magnetic response of C$_{60}$, which in bulk is even less diamagnetic than C$_{70}$. The magnetic response is consistent with atom-like deflection of rotational magnetic moments. Particularly noteworthy is the emergence of this quantized behavior at high rotational temperatures, where the classical limit is typically reached. The robustness of this behavior stems from the high symmetry of the system.

\begin{acknowledgments}
We would like to acknowledge the experimental assistance of Tom\'{a}s Sousa, Philipp Geyer and Richard Ferstl as well as the University of Vienna NMR Center for the C$_{60}$ analysis. This project has received funding from the European Research Council (ERC) under the European Union’s Horizon 2020 Research and Innovation Program (Grant No. 320694), and the Austrian Science Fund (FWF) within programs P-30176, W-1210-N25 and P32543-N. The computational results presented were obtained using the Vienna Scientific Cluster (VSC) within Grant No. 70918.
\end{acknowledgments}

\nocite{hornberger03,cheeseman96,gauss96,wilson05,kendall92} 

\bibliography{Fein2022}

\begin{thebibliography}{56}%
\makeatletter
\providecommand \@ifxundefined [1]{%
 \@ifx{#1\undefined}
}%
\providecommand \@ifnum [1]{%
 \ifnum #1\expandafter \@firstoftwo
 \else \expandafter \@secondoftwo
 \fi
}%
\providecommand \@ifx [1]{%
 \ifx #1\expandafter \@firstoftwo
 \else \expandafter \@secondoftwo
 \fi
}%
\providecommand \natexlab [1]{#1}%
\providecommand \enquote  [1]{``#1''}%
\providecommand \bibnamefont  [1]{#1}%
\providecommand \bibfnamefont [1]{#1}%
\providecommand \citenamefont [1]{#1}%
\providecommand \href@noop [0]{\@secondoftwo}%
\providecommand \href [0]{\begingroup \@sanitize@url \@href}%
\providecommand \@href[1]{\@@startlink{#1}\@@href}%
\providecommand \@@href[1]{\endgroup#1\@@endlink}%
\providecommand \@sanitize@url [0]{\catcode `\\12\catcode `\$12\catcode
  `\&12\catcode `\#12\catcode `\^12\catcode `\_12\catcode `\%12\relax}%
\providecommand \@@startlink[1]{}%
\providecommand \@@endlink[0]{}%
\providecommand \url  [0]{\begingroup\@sanitize@url \@url }%
\providecommand \@url [1]{\endgroup\@href {#1}{\urlprefix }}%
\providecommand \urlprefix  [0]{URL }%
\providecommand \Eprint [0]{\href }%
\providecommand \doibase [0]{https://doi.org/}%
\providecommand \selectlanguage [0]{\@gobble}%
\providecommand \bibinfo  [0]{\@secondoftwo}%
\providecommand \bibfield  [0]{\@secondoftwo}%
\providecommand \translation [1]{[#1]}%
\providecommand \BibitemOpen [0]{}%
\providecommand \bibitemStop [0]{}%
\providecommand \bibitemNoStop [0]{.\EOS\space}%
\providecommand \EOS [0]{\spacefactor3000\relax}%
\providecommand \BibitemShut  [1]{\csname bibitem#1\endcsname}%
\let\auto@bib@innerbib\@empty
\bibitem [{\citenamefont {{Gerlach}}\ and\ \citenamefont
  {{Stern}}(1922)}]{gerlach22}%
  \BibitemOpen
  \bibfield  {author} {\bibinfo {author} {\bibfnamefont {W.}~\bibnamefont
  {{Gerlach}}}\ and\ \bibinfo {author} {\bibfnamefont {O.}~\bibnamefont
  {{Stern}}},\ }\bibfield  {title} {\bibinfo {title} {{Der experimentelle
  Nachweis des magnetischen Moments des Silberatoms}},\ }\href
  {https://doi.org/10.1007/BF01329580} {\bibfield  {journal} {\bibinfo
  {journal} {Z. Phys.}\ }\textbf {\bibinfo {volume} {8}},\ \bibinfo {pages}
  {110} (\bibinfo {year} {1922})}\BibitemShut {NoStop}%
\bibitem [{\citenamefont {Van~Vleck}(1932)}]{vanVleck32}%
  \BibitemOpen
  \bibfield  {author} {\bibinfo {author} {\bibfnamefont {J.~H.}\ \bibnamefont
  {Van~Vleck}},\ }\href@noop {} {\emph {\bibinfo {title} {The theory of
  electric and magnetic susceptibilities}}}\ (\bibinfo  {publisher} {Oxford
  University Press},\ \bibinfo {year} {1932})\BibitemShut {NoStop}%
\bibitem [{\citenamefont {Allman}\ \emph {et~al.}(1993)\citenamefont {Allman},
  \citenamefont {Cimmino}, \citenamefont {Klein}, \citenamefont {Opat},
  \citenamefont {Kaiser},\ and\ \citenamefont {Werner}}]{allman93}%
  \BibitemOpen
  \bibfield  {author} {\bibinfo {author} {\bibfnamefont {B.~E.}\ \bibnamefont
  {Allman}}, \bibinfo {author} {\bibfnamefont {A.}~\bibnamefont {Cimmino}},
  \bibinfo {author} {\bibfnamefont {A.~G.}\ \bibnamefont {Klein}}, \bibinfo
  {author} {\bibfnamefont {G.~I.}\ \bibnamefont {Opat}}, \bibinfo {author}
  {\bibfnamefont {H.}~\bibnamefont {Kaiser}},\ and\ \bibinfo {author}
  {\bibfnamefont {S.~A.}\ \bibnamefont {Werner}},\ }\bibfield  {title}
  {\bibinfo {title} {Observation of the scalar {Aharonov-Bohm} effect by
  neutron interferometry},\ }\href {https://doi.org/10.1103/PhysRevA.48.1799}
  {\bibfield  {journal} {\bibinfo  {journal} {Phys. Rev. A}\ }\textbf {\bibinfo
  {volume} {48}},\ \bibinfo {pages} {1799} (\bibinfo {year}
  {1993})}\BibitemShut {NoStop}%
\bibitem [{\citenamefont {Wagh}\ \emph {et~al.}(1997)\citenamefont {Wagh},
  \citenamefont {Rakhecha}, \citenamefont {Summhammer}, \citenamefont
  {Badurek}, \citenamefont {Weinfurter}, \citenamefont {Allman}, \citenamefont
  {Kaiser}, \citenamefont {Hamacher}, \citenamefont {Jacobson},\ and\
  \citenamefont {Werner}}]{wagh97}%
  \BibitemOpen
  \bibfield  {author} {\bibinfo {author} {\bibfnamefont {A.~G.}\ \bibnamefont
  {Wagh}}, \bibinfo {author} {\bibfnamefont {V.~C.}\ \bibnamefont {Rakhecha}},
  \bibinfo {author} {\bibfnamefont {J.}~\bibnamefont {Summhammer}}, \bibinfo
  {author} {\bibfnamefont {G.}~\bibnamefont {Badurek}}, \bibinfo {author}
  {\bibfnamefont {H.}~\bibnamefont {Weinfurter}}, \bibinfo {author}
  {\bibfnamefont {B.~E.}\ \bibnamefont {Allman}}, \bibinfo {author}
  {\bibfnamefont {H.}~\bibnamefont {Kaiser}}, \bibinfo {author} {\bibfnamefont
  {K.}~\bibnamefont {Hamacher}}, \bibinfo {author} {\bibfnamefont {D.~L.}\
  \bibnamefont {Jacobson}},\ and\ \bibinfo {author} {\bibfnamefont {S.~A.}\
  \bibnamefont {Werner}},\ }\bibfield  {title} {\bibinfo {title} {Experimental
  separation of geometric and dynamical phases using neutron interferometry},\
  }\href {https://doi.org/10.1103/PhysRevLett.78.755} {\bibfield  {journal}
  {\bibinfo  {journal} {Phys. Rev. Lett.}\ }\textbf {\bibinfo {volume} {78}},\
  \bibinfo {pages} {755} (\bibinfo {year} {1997})}\BibitemShut {NoStop}%
\bibitem [{\citenamefont {Rauch}\ \emph {et~al.}(1975)\citenamefont {Rauch},
  \citenamefont {Zeilinger}, \citenamefont {Badurek}, \citenamefont {Wilfing},
  \citenamefont {Bauspiess},\ and\ \citenamefont {Bonse}}]{rauch75}%
  \BibitemOpen
  \bibfield  {author} {\bibinfo {author} {\bibfnamefont {H.}~\bibnamefont
  {Rauch}}, \bibinfo {author} {\bibfnamefont {A.}~\bibnamefont {Zeilinger}},
  \bibinfo {author} {\bibfnamefont {G.}~\bibnamefont {Badurek}}, \bibinfo
  {author} {\bibfnamefont {A.}~\bibnamefont {Wilfing}}, \bibinfo {author}
  {\bibfnamefont {W.}~\bibnamefont {Bauspiess}},\ and\ \bibinfo {author}
  {\bibfnamefont {U.}~\bibnamefont {Bonse}},\ }\bibfield  {title} {\bibinfo
  {title} {Verification of coherent spinor rotation of fermions},\ }\href
  {https://doi.org/https://doi.org/10.1016/0375-9601(75)90798-7} {\bibfield
  {journal} {\bibinfo  {journal} {Phys. Lett. A}\ }\textbf {\bibinfo {volume}
  {54}},\ \bibinfo {pages} {425} (\bibinfo {year} {1975})}\BibitemShut
  {NoStop}%
\bibitem [{\citenamefont {Werner}\ \emph {et~al.}(1975)\citenamefont {Werner},
  \citenamefont {Colella}, \citenamefont {Overhauser},\ and\ \citenamefont
  {Eagen}}]{werner75}%
  \BibitemOpen
  \bibfield  {author} {\bibinfo {author} {\bibfnamefont {S.~A.}\ \bibnamefont
  {Werner}}, \bibinfo {author} {\bibfnamefont {R.}~\bibnamefont {Colella}},
  \bibinfo {author} {\bibfnamefont {A.~W.}\ \bibnamefont {Overhauser}},\ and\
  \bibinfo {author} {\bibfnamefont {C.~F.}\ \bibnamefont {Eagen}},\ }\bibfield
  {title} {\bibinfo {title} {Observation of the phase shift of a neutron due to
  precession in a magnetic field},\ }\href
  {https://doi.org/10.1103/PhysRevLett.35.1053} {\bibfield  {journal} {\bibinfo
   {journal} {Phys. Rev. Lett.}\ }\textbf {\bibinfo {volume} {35}},\ \bibinfo
  {pages} {1053} (\bibinfo {year} {1975})}\BibitemShut {NoStop}%
\bibitem [{\citenamefont {Machluf}\ \emph {et~al.}(2013)\citenamefont
  {Machluf}, \citenamefont {Japha},\ and\ \citenamefont {Folman}}]{machluf13}%
  \BibitemOpen
  \bibfield  {author} {\bibinfo {author} {\bibfnamefont {S.}~\bibnamefont
  {Machluf}}, \bibinfo {author} {\bibfnamefont {Y.}~\bibnamefont {Japha}},\
  and\ \bibinfo {author} {\bibfnamefont {R.}~\bibnamefont {Folman}},\
  }\bibfield  {title} {\bibinfo {title} {Coherent {S}tern-{G}erlach momentum
  splitting on an atom chip},\ }\href {https://doi.org/10.1038/ncomms3424}
  {\bibfield  {journal} {\bibinfo  {journal} {Nature Commun.}\ }\textbf
  {\bibinfo {volume} {4}},\ \bibinfo {pages} {2424} (\bibinfo {year}
  {2013})}\BibitemShut {NoStop}%
\bibitem [{\citenamefont {Fein}\ \emph {et~al.}(2019)\citenamefont {Fein},
  \citenamefont {Geyer}, \citenamefont {Zwick}, \citenamefont {Kiałka},
  \citenamefont {Pedalino}, \citenamefont {Mayor}, \citenamefont {Gerlich},\
  and\ \citenamefont {Arndt}}]{fein19}%
  \BibitemOpen
  \bibfield  {author} {\bibinfo {author} {\bibfnamefont {Y.~Y.}\ \bibnamefont
  {Fein}}, \bibinfo {author} {\bibfnamefont {P.}~\bibnamefont {Geyer}},
  \bibinfo {author} {\bibfnamefont {P.}~\bibnamefont {Zwick}}, \bibinfo
  {author} {\bibfnamefont {F.}~\bibnamefont {Kiałka}}, \bibinfo {author}
  {\bibfnamefont {S.}~\bibnamefont {Pedalino}}, \bibinfo {author}
  {\bibfnamefont {M.}~\bibnamefont {Mayor}}, \bibinfo {author} {\bibfnamefont
  {S.}~\bibnamefont {Gerlich}},\ and\ \bibinfo {author} {\bibfnamefont
  {M.}~\bibnamefont {Arndt}},\ }\bibfield  {title} {\bibinfo {title} {Quantum
  superposition of molecules beyond 25 {kDa}},\ }\href
  {https://doi.org/10.1038/s41567-019-0663-9} {\bibfield  {journal} {\bibinfo
  {journal} {Nature Phys.}\ }\textbf {\bibinfo {volume} {15}},\ \bibinfo
  {pages} {1242} (\bibinfo {year} {2019})}\BibitemShut {NoStop}%
\bibitem [{\citenamefont {Fein}(2020)}]{fein20c}%
  \BibitemOpen
  \bibfield  {author} {\bibinfo {author} {\bibfnamefont {Y.~Y.}\ \bibnamefont
  {Fein}},\ }\emph {\bibinfo {title} {Long-baseline universal matter-wave
  interferometry}},\ \href {https://doi.org/10.25365/thesis.62746} {Ph.D.
  thesis},\ \bibinfo  {school} {University of Vienna} (\bibinfo {year}
  {2020})\BibitemShut {NoStop}%
\bibitem [{\citenamefont {Fein}\ \emph
  {et~al.}(2020{\natexlab{a}})\citenamefont {Fein}, \citenamefont {Shayeghi},
  \citenamefont {Mairhofer}, \citenamefont {Kia\l{}ka}, \citenamefont {Rieser},
  \citenamefont {Geyer}, \citenamefont {Gerlich},\ and\ \citenamefont
  {Arndt}}]{fein20}%
  \BibitemOpen
  \bibfield  {author} {\bibinfo {author} {\bibfnamefont {Y.~Y.}\ \bibnamefont
  {Fein}}, \bibinfo {author} {\bibfnamefont {A.}~\bibnamefont {Shayeghi}},
  \bibinfo {author} {\bibfnamefont {L.}~\bibnamefont {Mairhofer}}, \bibinfo
  {author} {\bibfnamefont {F.}~\bibnamefont {Kia\l{}ka}}, \bibinfo {author}
  {\bibfnamefont {P.}~\bibnamefont {Rieser}}, \bibinfo {author} {\bibfnamefont
  {P.}~\bibnamefont {Geyer}}, \bibinfo {author} {\bibfnamefont
  {S.}~\bibnamefont {Gerlich}},\ and\ \bibinfo {author} {\bibfnamefont
  {M.}~\bibnamefont {Arndt}},\ }\bibfield  {title} {\bibinfo {title}
  {Quantum-assisted measurement of atomic diamagnetism},\ }\href
  {https://doi.org/10.1103/PhysRevX.10.011014} {\bibfield  {journal} {\bibinfo
  {journal} {Phys. Rev. X}\ }\textbf {\bibinfo {volume} {10}},\ \bibinfo
  {pages} {011014} (\bibinfo {year} {2020}{\natexlab{a}})}\BibitemShut
  {NoStop}%
\bibitem [{\citenamefont {Gerlich}\ \emph {et~al.}(2021)\citenamefont
  {Gerlich}, \citenamefont {Fein}, \citenamefont {Shayeghi}, \citenamefont
  {K{\"o}hler}, \citenamefont {Mayor},\ and\ \citenamefont
  {Arndt}}]{gerlich21}%
  \BibitemOpen
  \bibfield  {author} {\bibinfo {author} {\bibfnamefont {S.}~\bibnamefont
  {Gerlich}}, \bibinfo {author} {\bibfnamefont {Y.~Y.}\ \bibnamefont {Fein}},
  \bibinfo {author} {\bibfnamefont {A.}~\bibnamefont {Shayeghi}}, \bibinfo
  {author} {\bibfnamefont {V.}~\bibnamefont {K{\"o}hler}}, \bibinfo {author}
  {\bibfnamefont {M.}~\bibnamefont {Mayor}},\ and\ \bibinfo {author}
  {\bibfnamefont {M.}~\bibnamefont {Arndt}},\ }\bibinfo {title} {Otto {Stern's}
  legacy in quantum optics: Matter waves and deflectometry},\ in\ \href
  {https://doi.org/10.1007/978-3-030-63963-1_24} {\emph {\bibinfo {booktitle}
  {Molecular Beams in Physics and Chemistry: From Otto Stern's Pioneering
  Exploits to Present-Day Feats}}},\ \bibinfo {editor} {edited by\ \bibinfo
  {editor} {\bibfnamefont {B.}~\bibnamefont {Friedrich}}\ and\ \bibinfo
  {editor} {\bibfnamefont {H.}~\bibnamefont {Schmidt-B{\"o}cking}}}\ (\bibinfo
  {publisher} {Springer International Publishing},\ \bibinfo {year} {2021})\
  pp.\ \bibinfo {pages} {547--573}\BibitemShut {NoStop}%
\bibitem [{\citenamefont {Hornberger}\ \emph {et~al.}(2012)\citenamefont
  {Hornberger}, \citenamefont {Gerlich}, \citenamefont {Haslinger},
  \citenamefont {Nimmrichter},\ and\ \citenamefont {Arndt}}]{hornberger12}%
  \BibitemOpen
  \bibfield  {author} {\bibinfo {author} {\bibfnamefont {K.}~\bibnamefont
  {Hornberger}}, \bibinfo {author} {\bibfnamefont {S.}~\bibnamefont {Gerlich}},
  \bibinfo {author} {\bibfnamefont {P.}~\bibnamefont {Haslinger}}, \bibinfo
  {author} {\bibfnamefont {S.}~\bibnamefont {Nimmrichter}},\ and\ \bibinfo
  {author} {\bibfnamefont {M.}~\bibnamefont {Arndt}},\ }\bibfield  {title}
  {\bibinfo {title} {Colloquium: Quantum interference of clusters and
  molecules},\ }\href {https://doi.org/10.1103/RevModPhys.84.157} {\bibfield
  {journal} {\bibinfo  {journal} {Rev. Mod. Phys.}\ }\textbf {\bibinfo {volume}
  {84}},\ \bibinfo {pages} {157} (\bibinfo {year} {2012})}\BibitemShut
  {NoStop}%
\bibitem [{\citenamefont {Clauser}\ and\ \citenamefont {Li}(1997)}]{clauser97}%
  \BibitemOpen
  \bibfield  {author} {\bibinfo {author} {\bibfnamefont {J.~F.}\ \bibnamefont
  {Clauser}}\ and\ \bibinfo {author} {\bibfnamefont {S.}~\bibnamefont {Li}},\
  }\bibfield  {title} {\bibinfo {title} {Generalized {Talbot-Lau} atom
  interferometry},\ }in\ \href
  {https://doi.org/https://doi.org/10.1016/B978-012092460-8/50004-7} {\emph
  {\bibinfo {booktitle} {Atom Interferometry}}},\ \bibinfo {editor} {edited by\
  \bibinfo {editor} {\bibfnamefont {P.~R.}\ \bibnamefont {Berman}}}\ (\bibinfo
  {publisher} {Academic Press},\ \bibinfo {address} {San Diego},\ \bibinfo
  {year} {1997})\ pp.\ \bibinfo {pages} {121--151}\BibitemShut {NoStop}%
\bibitem [{\citenamefont {Schmiedmayer}\ \emph {et~al.}(1994)\citenamefont
  {Schmiedmayer}, \citenamefont {Ekstrom}, \citenamefont {Chapman},
  \citenamefont {Hammond},\ and\ \citenamefont {Pritchard}}]{schmiedmayer94}%
  \BibitemOpen
  \bibfield  {author} {\bibinfo {author} {\bibfnamefont {J.}~\bibnamefont
  {Schmiedmayer}}, \bibinfo {author} {\bibfnamefont {C.}~\bibnamefont
  {Ekstrom}}, \bibinfo {author} {\bibfnamefont {M.}~\bibnamefont {Chapman}},
  \bibinfo {author} {\bibfnamefont {T.}~\bibnamefont {Hammond}},\ and\ \bibinfo
  {author} {\bibfnamefont {D.}~\bibnamefont {Pritchard}},\ }\bibfield  {title}
  {\bibinfo {title} {Magnetic coherences in atom interferometry},\ }\href
  {https://doi.org/10.1051/jp2:1994245} {\bibfield  {journal} {\bibinfo
  {journal} {J. Phys. II}\ }\textbf {\bibinfo {volume} {4}},\ \bibinfo {pages}
  {2029} (\bibinfo {year} {1994})}\BibitemShut {NoStop}%
\bibitem [{\citenamefont {Jacquey}\ \emph {et~al.}(2007)\citenamefont
  {Jacquey}, \citenamefont {Miffre}, \citenamefont {Büchner}, \citenamefont
  {Tr{\'{e}}nec},\ and\ \citenamefont {Vigu{\'{e}}}}]{Jacquey07}%
  \BibitemOpen
  \bibfield  {author} {\bibinfo {author} {\bibfnamefont {M.}~\bibnamefont
  {Jacquey}}, \bibinfo {author} {\bibfnamefont {A.}~\bibnamefont {Miffre}},
  \bibinfo {author} {\bibfnamefont {M.}~\bibnamefont {Büchner}}, \bibinfo
  {author} {\bibfnamefont {G.}~\bibnamefont {Tr{\'{e}}nec}},\ and\ \bibinfo
  {author} {\bibfnamefont {J.}~\bibnamefont {Vigu{\'{e}}}},\ }\bibfield
  {title} {\bibinfo {title} {Test of the isotopic and velocity selectivity of a
  lithium atom interferometer by magnetic dephasing},\ }\href
  {https://doi.org/10.1209/0295-5075/77/20007} {\bibfield  {journal} {\bibinfo
  {journal} {Europhysics Letters ({EPL})}\ }\textbf {\bibinfo {volume} {77}},\
  \bibinfo {pages} {20007} (\bibinfo {year} {2007})}\BibitemShut {NoStop}%
\bibitem [{\citenamefont {de~Heer}\ and\ \citenamefont
  {Kresin}(2010)}]{deHeer10}%
  \BibitemOpen
  \bibfield  {author} {\bibinfo {author} {\bibfnamefont {W.~A.}\ \bibnamefont
  {de~Heer}}\ and\ \bibinfo {author} {\bibfnamefont {V.~V.}\ \bibnamefont
  {Kresin}},\ }\bibfield  {title} {\bibinfo {title} {The electronic structure
  of alkali and noble metal clusters},\ }in\ \href
  {https://doi.org/10.1201/9781420075557-12} {\emph {\bibinfo {booktitle}
  {Handbook of Nanophysics}}}\ (\bibinfo  {publisher} {{CRC} Press},\ \bibinfo
  {year} {2010})\ Chap.\ \bibinfo {chapter} {Electric and Magnetic Dipole
  Moments of Free Nanoclusters}, pp.\ \bibinfo {pages} {97--116}\BibitemShut
  {NoStop}%
\bibitem [{\citenamefont {Khanna}\ and\ \citenamefont
  {Linderoth}(1991)}]{khanna91}%
  \BibitemOpen
  \bibfield  {author} {\bibinfo {author} {\bibfnamefont {S.~N.}\ \bibnamefont
  {Khanna}}\ and\ \bibinfo {author} {\bibfnamefont {S.}~\bibnamefont
  {Linderoth}},\ }\bibfield  {title} {\bibinfo {title} {Magnetic behavior of
  clusters of ferromagnetic transition metals},\ }\href
  {https://doi.org/10.1103/PhysRevLett.67.742} {\bibfield  {journal} {\bibinfo
  {journal} {Phys. Rev. Lett.}\ }\textbf {\bibinfo {volume} {67}},\ \bibinfo
  {pages} {742} (\bibinfo {year} {1991})}\BibitemShut {NoStop}%
\bibitem [{\citenamefont {Xu}\ \emph {et~al.}(2005)\citenamefont {Xu},
  \citenamefont {Yin}, \citenamefont {Moro},\ and\ \citenamefont
  {de~Heer}}]{xu05}%
  \BibitemOpen
  \bibfield  {author} {\bibinfo {author} {\bibfnamefont {X.}~\bibnamefont
  {Xu}}, \bibinfo {author} {\bibfnamefont {S.}~\bibnamefont {Yin}}, \bibinfo
  {author} {\bibfnamefont {R.}~\bibnamefont {Moro}},\ and\ \bibinfo {author}
  {\bibfnamefont {W.~A.}\ \bibnamefont {de~Heer}},\ }\bibfield  {title}
  {\bibinfo {title} {Magnetic moments and adiabatic magnetization of free
  cobalt clusters},\ }\href {https://doi.org/10.1103/PhysRevLett.95.237209}
  {\bibfield  {journal} {\bibinfo  {journal} {Phys. Rev. Lett.}\ }\textbf
  {\bibinfo {volume} {95}},\ \bibinfo {pages} {237209} (\bibinfo {year}
  {2005})}\BibitemShut {NoStop}%
\bibitem [{\citenamefont {Knickelbein}(2004)}]{knickelbein04}%
  \BibitemOpen
  \bibfield  {author} {\bibinfo {author} {\bibfnamefont {M.~B.}\ \bibnamefont
  {Knickelbein}},\ }\bibfield  {title} {\bibinfo {title} {Spin relaxation in
  isolated molecules and clusters:{\hspace{1em}}the interpretation of
  {Stern-Gerlach} experiments},\ }\href {https://doi.org/10.1063/1.1781156}
  {\bibfield  {journal} {\bibinfo  {journal} {The Journal of Chemical Physics}\
  }\textbf {\bibinfo {volume} {121}},\ \bibinfo {pages} {5281 } (\bibinfo
  {year} {2004})}\BibitemShut {NoStop}%
\bibitem [{\citenamefont {Coronado}(2019)}]{coronado19}%
  \BibitemOpen
  \bibfield  {author} {\bibinfo {author} {\bibfnamefont {E.}~\bibnamefont
  {Coronado}},\ }\bibfield  {title} {\bibinfo {title} {Molecular magnetism:
  from chemical design to spin control in molecules, materials and devices},\
  }\href {https://doi.org/10.1038/s41578-019-0146-8} {\bibfield  {journal}
  {\bibinfo  {journal} {Nature Reviews Materials}\ }\textbf {\bibinfo {volume}
  {5}},\ \bibinfo {pages} {87} (\bibinfo {year} {2019})}\BibitemShut {NoStop}%
\bibitem [{\citenamefont {Gerlich}\ \emph {et~al.}(2008)\citenamefont
  {Gerlich}, \citenamefont {Hackermueller}, \citenamefont {Hornberger},
  \citenamefont {Stibor}, \citenamefont {Ulbricht}, \citenamefont {Gring},
  \citenamefont {Goldfarb}, \citenamefont {Savas}, \citenamefont {Mueri},
  \citenamefont {Mayor},\ and\ \citenamefont {Arndt}}]{gerlich08}%
  \BibitemOpen
  \bibfield  {author} {\bibinfo {author} {\bibfnamefont {S.}~\bibnamefont
  {Gerlich}}, \bibinfo {author} {\bibfnamefont {L.}~\bibnamefont
  {Hackermueller}}, \bibinfo {author} {\bibfnamefont {K.}~\bibnamefont
  {Hornberger}}, \bibinfo {author} {\bibfnamefont {A.}~\bibnamefont {Stibor}},
  \bibinfo {author} {\bibfnamefont {H.}~\bibnamefont {Ulbricht}}, \bibinfo
  {author} {\bibfnamefont {M.}~\bibnamefont {Gring}}, \bibinfo {author}
  {\bibfnamefont {F.}~\bibnamefont {Goldfarb}}, \bibinfo {author}
  {\bibfnamefont {T.}~\bibnamefont {Savas}}, \bibinfo {author} {\bibfnamefont
  {M.}~\bibnamefont {Mueri}}, \bibinfo {author} {\bibfnamefont
  {M.}~\bibnamefont {Mayor}},\ and\ \bibinfo {author} {\bibfnamefont
  {M.}~\bibnamefont {Arndt}},\ }\bibfield  {title} {\bibinfo {title} {A
  {Kapitza-Dirac-Talbot-Lau} interferometer for highly polarizable molecules},\
  }\href {https://doi.org/10.1038/nphys701} {\bibfield  {journal} {\bibinfo
  {journal} {Nature Phys.}\ }\textbf {\bibinfo {volume} {3}} (\bibinfo {year}
  {2008})}\BibitemShut {NoStop}%
\bibitem [{Note1()}]{Note1}%
  \BibitemOpen
  \bibinfo {note} {See Supplemental Material, which includes Refs.~\cite{hornberger03,cheeseman96,gauss96,wilson05,kendall92}, for details on the characterization and modeling of the gradient
  field regions, additional experimental and data analysis details, and
  computational details for the rotational magnetic moments.}\BibitemShut
  {Stop}%
\bibitem [{\citenamefont {Bergeman}\ \emph {et~al.}(1987)\citenamefont
  {Bergeman}, \citenamefont {Erez},\ and\ \citenamefont
  {Metcalf}}]{bergeman87}%
  \BibitemOpen
  \bibfield  {author} {\bibinfo {author} {\bibfnamefont {T.}~\bibnamefont
  {Bergeman}}, \bibinfo {author} {\bibfnamefont {G.}~\bibnamefont {Erez}},\
  and\ \bibinfo {author} {\bibfnamefont {H.~J.}\ \bibnamefont {Metcalf}},\
  }\bibfield  {title} {\bibinfo {title} {Magnetostatic trapping fields for
  neutral atoms},\ }\href {https://doi.org/10.1103/PhysRevA.35.1535} {\bibfield
   {journal} {\bibinfo  {journal} {Phys. Rev. A}\ }\textbf {\bibinfo {volume}
  {35}},\ \bibinfo {pages} {1535} (\bibinfo {year} {1987})}\BibitemShut
  {NoStop}%
\bibitem [{\citenamefont {Camacho}\ and\ \citenamefont
  {Sosa}(2013)}]{camacho13}%
  \BibitemOpen
  \bibfield  {author} {\bibinfo {author} {\bibfnamefont {J.~M.}\ \bibnamefont
  {Camacho}}\ and\ \bibinfo {author} {\bibfnamefont {V.}~\bibnamefont {Sosa}},\
  }\bibfield  {title} {\bibinfo {title} {Alternative method to calculate the
  magnetic field of permanent magnets with azimuthal symmetry},\ }\href@noop {}
  {\bibfield  {journal} {\bibinfo  {journal} {Revista mexicana de f{\'\i}sica
  E}\ }\textbf {\bibinfo {volume} {59}},\ \bibinfo {pages} {8} (\bibinfo {year}
  {2013})}\BibitemShut {NoStop}%
\bibitem [{Note2()}]{Note2}%
  \BibitemOpen
  \bibinfo {note} {While the magnetic force itself is essentially classical
  (i.e., a phase gradient causing an envelope phase shift of the interference
  pattern) and the semi-classical formalism is used throughout, one obtains
  identical results for the phase shift with a path integral approach~\cite
  {Storey94}.}\BibitemShut {Stop}%
\bibitem [{\citenamefont {Cronin}\ \emph {et~al.}(2009)\citenamefont {Cronin},
  \citenamefont {Schmiedmayer},\ and\ \citenamefont {Pritchard}}]{cronin09}%
  \BibitemOpen
  \bibfield  {author} {\bibinfo {author} {\bibfnamefont {A.~D.}\ \bibnamefont
  {Cronin}}, \bibinfo {author} {\bibfnamefont {J.}~\bibnamefont
  {Schmiedmayer}},\ and\ \bibinfo {author} {\bibfnamefont {D.~E.}\ \bibnamefont
  {Pritchard}},\ }\bibfield  {title} {\bibinfo {title} {Optics and
  interferometry with atoms and molecules},\ }\href
  {https://doi.org/10.1103/RevModPhys.81.1051} {\bibfield  {journal} {\bibinfo
  {journal} {Rev. Mod. Phys.}\ }\textbf {\bibinfo {volume} {81}},\ \bibinfo
  {pages} {1051} (\bibinfo {year} {2009})}\BibitemShut {NoStop}%
\bibitem [{\citenamefont {Nimmrichter}\ and\ \citenamefont
  {Hornberger}(2008)}]{nimmrichter08}%
  \BibitemOpen
  \bibfield  {author} {\bibinfo {author} {\bibfnamefont {S.}~\bibnamefont
  {Nimmrichter}}\ and\ \bibinfo {author} {\bibfnamefont {K.}~\bibnamefont
  {Hornberger}},\ }\bibfield  {title} {\bibinfo {title} {Theory of near-field
  matter-wave interference beyond the eikonal approximation},\ }\href
  {https://doi.org/10.1103/PhysRevA.78.023612} {\bibfield  {journal} {\bibinfo
  {journal} {Phys. Rev. A}\ }\textbf {\bibinfo {volume} {78}},\ \bibinfo
  {pages} {023612} (\bibinfo {year} {2008})}\BibitemShut {NoStop}%
\bibitem [{\citenamefont {Mairhofer}\ \emph {et~al.}(2018)\citenamefont
  {Mairhofer}, \citenamefont {Eibenberger}, \citenamefont {Shayeghi},\ and\
  \citenamefont {Arndt}}]{mairhofer18}%
  \BibitemOpen
  \bibfield  {author} {\bibinfo {author} {\bibfnamefont {L.}~\bibnamefont
  {Mairhofer}}, \bibinfo {author} {\bibfnamefont {S.}~\bibnamefont
  {Eibenberger}}, \bibinfo {author} {\bibfnamefont {A.}~\bibnamefont
  {Shayeghi}},\ and\ \bibinfo {author} {\bibfnamefont {M.}~\bibnamefont
  {Arndt}},\ }\bibfield  {title} {\bibinfo {title} {A quantum ruler for
  magnetic deflectometry},\ }\href {https://www.mdpi.com/1099-4300/20/7/516}
  {\bibfield  {journal} {\bibinfo  {journal} {Entropy}\ }\textbf {\bibinfo
  {volume} {20}} (\bibinfo {year} {2018})}\BibitemShut {NoStop}%
\bibitem [{\citenamefont {Fein}\ \emph
  {et~al.}(2020{\natexlab{b}})\citenamefont {Fein}, \citenamefont {Shayeghi},
  \citenamefont {Kiałka}, \citenamefont {Geyer}, \citenamefont {Gerlich},\
  and\ \citenamefont {Arndt}}]{fein20b}%
  \BibitemOpen
  \bibfield  {author} {\bibinfo {author} {\bibfnamefont {Y.~Y.}\ \bibnamefont
  {Fein}}, \bibinfo {author} {\bibfnamefont {A.}~\bibnamefont {Shayeghi}},
  \bibinfo {author} {\bibfnamefont {F.}~\bibnamefont {Kiałka}}, \bibinfo
  {author} {\bibfnamefont {P.}~\bibnamefont {Geyer}}, \bibinfo {author}
  {\bibfnamefont {S.}~\bibnamefont {Gerlich}},\ and\ \bibinfo {author}
  {\bibfnamefont {M.}~\bibnamefont {Arndt}},\ }\bibfield  {title} {\bibinfo
  {title} {Quantum-assisted diamagnetic deflection of molecules},\ }\href
  {https://doi.org/10.1039/D0CP02211J} {\bibfield  {journal} {\bibinfo
  {journal} {Phys. Chem. Chem. Phys.}\ }\textbf {\bibinfo {volume} {22}},\
  \bibinfo {pages} {14036} (\bibinfo {year} {2020}{\natexlab{b}})}\BibitemShut
  {NoStop}%
\bibitem [{\citenamefont {Flygare}(1974)}]{flygare74}%
  \BibitemOpen
  \bibfield  {author} {\bibinfo {author} {\bibfnamefont {W.~H.}\ \bibnamefont
  {Flygare}},\ }\bibfield  {title} {\bibinfo {title} {Magnetic interactions in
  molecules and an analysis of molecular electronic charge distribution from
  magnetic parameters},\ }\href {https://doi.org/10.1021/cr60292a003}
  {\bibfield  {journal} {\bibinfo  {journal} {Chemical Reviews}\ }\textbf
  {\bibinfo {volume} {74}},\ \bibinfo {pages} {653} (\bibinfo {year}
  {1974})}\BibitemShut {NoStop}%
\bibitem [{\citenamefont {Cox}\ \emph {et~al.}(1985)\citenamefont {Cox},
  \citenamefont {Trevor}, \citenamefont {Whetten}, \citenamefont {Rohlfing},\
  and\ \citenamefont {Kaldor}}]{cox85}%
  \BibitemOpen
  \bibfield  {author} {\bibinfo {author} {\bibfnamefont {D.~M.}\ \bibnamefont
  {Cox}}, \bibinfo {author} {\bibfnamefont {D.~J.}\ \bibnamefont {Trevor}},
  \bibinfo {author} {\bibfnamefont {R.~L.}\ \bibnamefont {Whetten}}, \bibinfo
  {author} {\bibfnamefont {E.~A.}\ \bibnamefont {Rohlfing}},\ and\ \bibinfo
  {author} {\bibfnamefont {A.}~\bibnamefont {Kaldor}},\ }\bibfield  {title}
  {\bibinfo {title} {Magnetic behavior of free-iron and iron oxide clusters},\
  }\href {https://doi.org/10.1103/PhysRevB.32.7290} {\bibfield  {journal}
  {\bibinfo  {journal} {Phys. Rev. B}\ }\textbf {\bibinfo {volume} {32}},\
  \bibinfo {pages} {7290} (\bibinfo {year} {1985})}\BibitemShut {NoStop}%
\bibitem [{\citenamefont {de~Heer}\ \emph {et~al.}(1990)\citenamefont
  {de~Heer}, \citenamefont {Milani},\ and\ \citenamefont
  {Chatelain}}]{deHeer90}%
  \BibitemOpen
  \bibfield  {author} {\bibinfo {author} {\bibfnamefont {W.~A.}\ \bibnamefont
  {de~Heer}}, \bibinfo {author} {\bibfnamefont {P.}~\bibnamefont {Milani}},\
  and\ \bibinfo {author} {\bibfnamefont {A.}~\bibnamefont {Chatelain}},\
  }\bibfield  {title} {\bibinfo {title} {Spin relaxation in small free iron
  clusters},\ }\href {https://doi.org/10.1103/PhysRevLett.65.488} {\bibfield
  {journal} {\bibinfo  {journal} {Phys. Rev. Lett.}\ }\textbf {\bibinfo
  {volume} {65}},\ \bibinfo {pages} {488} (\bibinfo {year} {1990})}\BibitemShut
  {NoStop}%
\bibitem [{\citenamefont {Bucher}\ \emph {et~al.}(1991)\citenamefont {Bucher},
  \citenamefont {Douglass},\ and\ \citenamefont {Bloomfield}}]{bucher91}%
  \BibitemOpen
  \bibfield  {author} {\bibinfo {author} {\bibfnamefont {J.~P.}\ \bibnamefont
  {Bucher}}, \bibinfo {author} {\bibfnamefont {D.~C.}\ \bibnamefont
  {Douglass}},\ and\ \bibinfo {author} {\bibfnamefont {L.~A.}\ \bibnamefont
  {Bloomfield}},\ }\bibfield  {title} {\bibinfo {title} {Magnetic properties of
  free cobalt clusters},\ }\href {https://doi.org/10.1103/PhysRevLett.66.3052}
  {\bibfield  {journal} {\bibinfo  {journal} {Phys. Rev. Lett.}\ }\textbf
  {\bibinfo {volume} {66}},\ \bibinfo {pages} {3052} (\bibinfo {year}
  {1991})}\BibitemShut {NoStop}%
\bibitem [{\citenamefont {Payne}\ \emph {et~al.}(2007)\citenamefont {Payne},
  \citenamefont {Jiang}, \citenamefont {Emmert}, \citenamefont {Deng},\ and\
  \citenamefont {Bloomfield}}]{payne07}%
  \BibitemOpen
  \bibfield  {author} {\bibinfo {author} {\bibfnamefont {F.~W.}\ \bibnamefont
  {Payne}}, \bibinfo {author} {\bibfnamefont {W.}~\bibnamefont {Jiang}},
  \bibinfo {author} {\bibfnamefont {J.~W.}\ \bibnamefont {Emmert}}, \bibinfo
  {author} {\bibfnamefont {J.}~\bibnamefont {Deng}},\ and\ \bibinfo {author}
  {\bibfnamefont {L.~A.}\ \bibnamefont {Bloomfield}},\ }\bibfield  {title}
  {\bibinfo {title} {Magnetic structure of free cobalt clusters studied with
  {Stern-Gerlach} deflection experiments},\ }\href
  {https://doi.org/10.1103/PhysRevB.75.094431} {\bibfield  {journal} {\bibinfo
  {journal} {Phys. Rev. B}\ }\textbf {\bibinfo {volume} {75}},\ \bibinfo
  {pages} {094431} (\bibinfo {year} {2007})}\BibitemShut {NoStop}%
\bibitem [{\citenamefont {Knickelbein}(2001)}]{knickelbein01}%
  \BibitemOpen
  \bibfield  {author} {\bibinfo {author} {\bibfnamefont {M.~B.}\ \bibnamefont
  {Knickelbein}},\ }\bibfield  {title} {\bibinfo {title} {Experimental
  observation of superparamagnetism in manganese clusters},\ }\href
  {https://doi.org/10.1103/PhysRevLett.86.5255} {\bibfield  {journal} {\bibinfo
   {journal} {Phys. Rev. Lett.}\ }\textbf {\bibinfo {volume} {86}},\ \bibinfo
  {pages} {5255} (\bibinfo {year} {2001})}\BibitemShut {NoStop}%
\bibitem [{\citenamefont {Xu}\ \emph {et~al.}(2008)\citenamefont {Xu},
  \citenamefont {Yin}, \citenamefont {Moro},\ and\ \citenamefont
  {de~Heer}}]{xu08}%
  \BibitemOpen
  \bibfield  {author} {\bibinfo {author} {\bibfnamefont {X.}~\bibnamefont
  {Xu}}, \bibinfo {author} {\bibfnamefont {S.}~\bibnamefont {Yin}}, \bibinfo
  {author} {\bibfnamefont {R.}~\bibnamefont {Moro}},\ and\ \bibinfo {author}
  {\bibfnamefont {W.~A.}\ \bibnamefont {de~Heer}},\ }\bibfield  {title}
  {\bibinfo {title} {Distribution of magnetization of a cold ferromagnetic
  cluster beam},\ }\href {https://doi.org/10.1103/PhysRevB.78.054430}
  {\bibfield  {journal} {\bibinfo  {journal} {Phys. Rev. B}\ }\textbf {\bibinfo
  {volume} {78}},\ \bibinfo {pages} {054430} (\bibinfo {year}
  {2008})}\BibitemShut {NoStop}%
\bibitem [{\citenamefont {Rohrmann}\ and\ \citenamefont
  {Sch\"afer}(2013)}]{rohrmann13}%
  \BibitemOpen
  \bibfield  {author} {\bibinfo {author} {\bibfnamefont {U.}~\bibnamefont
  {Rohrmann}}\ and\ \bibinfo {author} {\bibfnamefont {R.}~\bibnamefont
  {Sch\"afer}},\ }\bibfield  {title} {\bibinfo {title} {{S}tern-{G}erlach
  experiments on {$\mathrm{Mn}@{\mathrm{Sn}}_{12}$}: Identification of a
  paramagnetic superatom and vibrationally induced spin orientation},\ }\href
  {https://doi.org/10.1103/PhysRevLett.111.133401} {\bibfield  {journal}
  {\bibinfo  {journal} {Phys. Rev. Lett.}\ }\textbf {\bibinfo {volume} {111}},\
  \bibinfo {pages} {133401} (\bibinfo {year} {2013})}\BibitemShut {NoStop}%
\bibitem [{\citenamefont {Amirav}\ and\ \citenamefont
  {Navon}(1983)}]{amirav83}%
  \BibitemOpen
  \bibfield  {author} {\bibinfo {author} {\bibfnamefont {A.}~\bibnamefont
  {Amirav}}\ and\ \bibinfo {author} {\bibfnamefont {G.}~\bibnamefont {Navon}},\
  }\bibfield  {title} {\bibinfo {title} {Intramolecular spin relaxation probed
  by {Stern-Gerlach} experiments},\ }\href
  {https://doi.org/https://doi.org/10.1016/0301-0104(83)85233-1} {\bibfield
  {journal} {\bibinfo  {journal} {Chemical Physics}\ }\textbf {\bibinfo
  {volume} {82}},\ \bibinfo {pages} {253} (\bibinfo {year} {1983})}\BibitemShut
  {NoStop}%
\bibitem [{\citenamefont {Gedanken}\ \emph {et~al.}(1989)\citenamefont
  {Gedanken}, \citenamefont {Kuebler}, \citenamefont {Robin},\ and\
  \citenamefont {Herrick}}]{gedanken89}%
  \BibitemOpen
  \bibfield  {author} {\bibinfo {author} {\bibfnamefont {A.}~\bibnamefont
  {Gedanken}}, \bibinfo {author} {\bibfnamefont {N.~A.}\ \bibnamefont
  {Kuebler}}, \bibinfo {author} {\bibfnamefont {M.~B.}\ \bibnamefont {Robin}},\
  and\ \bibinfo {author} {\bibfnamefont {D.~R.}\ \bibnamefont {Herrick}},\
  }\bibfield  {title} {\bibinfo {title} {{Stern–Gerlach} deflection spectra
  of nitrogen oxide radicals},\ }\href {https://doi.org/10.1063/1.455808}
  {\bibfield  {journal} {\bibinfo  {journal} {The Journal of Chemical Physics}\
  }\textbf {\bibinfo {volume} {90}},\ \bibinfo {pages} {3981} (\bibinfo {year}
  {1989})}\BibitemShut {NoStop}%
\bibitem [{\citenamefont {Even}\ \emph {et~al.}(2000)\citenamefont {Even},
  \citenamefont {Jortner}, \citenamefont {Noy}, \citenamefont {Lavie},\ and\
  \citenamefont {Cossart-Magos}}]{even00}%
  \BibitemOpen
  \bibfield  {author} {\bibinfo {author} {\bibfnamefont {U.}~\bibnamefont
  {Even}}, \bibinfo {author} {\bibfnamefont {J.}~\bibnamefont {Jortner}},
  \bibinfo {author} {\bibfnamefont {D.}~\bibnamefont {Noy}}, \bibinfo {author}
  {\bibfnamefont {N.}~\bibnamefont {Lavie}},\ and\ \bibinfo {author}
  {\bibfnamefont {C.}~\bibnamefont {Cossart-Magos}},\ }\bibfield  {title}
  {\bibinfo {title} {Cooling of large molecules below {1 K and He} clusters
  formation},\ }\href@noop {} {\bibfield  {journal} {\bibinfo  {journal} {The
  Journal of Chemical Physics}\ }\textbf {\bibinfo {volume} {112}},\ \bibinfo
  {pages} {8068} (\bibinfo {year} {2000})}\BibitemShut {NoStop}%
\bibitem [{\citenamefont {Haddon}(1995)}]{Haddon95}%
  \BibitemOpen
  \bibfield  {author} {\bibinfo {author} {\bibfnamefont {R.~C.}\ \bibnamefont
  {Haddon}},\ }\bibfield  {title} {\bibinfo {title} {Magnetism of the carbon
  allotropes},\ }\href {https://doi.org/10.1038/378249a0} {\bibfield  {journal}
  {\bibinfo  {journal} {Nature}\ }\textbf {\bibinfo {volume} {378}},\ \bibinfo
  {pages} {249} (\bibinfo {year} {1995})}\BibitemShut {NoStop}%
\bibitem [{\citenamefont {Kim}\ \emph {et~al.}(2003)\citenamefont {Kim},
  \citenamefont {Choi}, \citenamefont {Chang},\ and\ \citenamefont
  {Tom\'anek}}]{kim03}%
  \BibitemOpen
  \bibfield  {author} {\bibinfo {author} {\bibfnamefont {Y.-H.}\ \bibnamefont
  {Kim}}, \bibinfo {author} {\bibfnamefont {J.}~\bibnamefont {Choi}}, \bibinfo
  {author} {\bibfnamefont {K.~J.}\ \bibnamefont {Chang}},\ and\ \bibinfo
  {author} {\bibfnamefont {D.}~\bibnamefont {Tom\'anek}},\ }\bibfield  {title}
  {\bibinfo {title} {Defective fullerenes and nanotubes as molecular magnets:
  An ab initio study},\ }\href {https://doi.org/10.1103/PhysRevB.68.125420}
  {\bibfield  {journal} {\bibinfo  {journal} {Phys. Rev. B}\ }\textbf {\bibinfo
  {volume} {68}},\ \bibinfo {pages} {125420} (\bibinfo {year}
  {2003})}\BibitemShut {NoStop}%
\bibitem [{\citenamefont {Eshbach}\ and\ \citenamefont
  {Strandberg}(1952)}]{eshbach52}%
  \BibitemOpen
  \bibfield  {author} {\bibinfo {author} {\bibfnamefont {J.~R.}\ \bibnamefont
  {Eshbach}}\ and\ \bibinfo {author} {\bibfnamefont {M.~W.~P.}\ \bibnamefont
  {Strandberg}},\ }\bibfield  {title} {\bibinfo {title} {Rotational magnetic
  moments of $^{1}\ensuremath{\Sigma}$ molecules},\ }\href
  {https://doi.org/10.1103/PhysRev.85.24} {\bibfield  {journal} {\bibinfo
  {journal} {Phys. Rev.}\ }\textbf {\bibinfo {volume} {85}},\ \bibinfo {pages}
  {24} (\bibinfo {year} {1952})}\BibitemShut {NoStop}%
\bibitem [{\citenamefont {Becke}(1993)}]{becke93}%
  \BibitemOpen
  \bibfield  {author} {\bibinfo {author} {\bibfnamefont {A.~D.}\ \bibnamefont
  {Becke}},\ }\bibfield  {title} {\bibinfo {title} {Density‐functional
  thermochemistry. {III}. {T}he role of exact exchange},\ }\href
  {https://doi.org/10.1063/1.464913} {\bibfield  {journal} {\bibinfo  {journal}
  {The Journal of Chemical Physics}\ }\textbf {\bibinfo {volume} {98}},\
  \bibinfo {pages} {5648} (\bibinfo {year} {1993})}\BibitemShut {NoStop}%
\bibitem [{\citenamefont {Weigend}\ and\ \citenamefont
  {Ahlrichs}(2005)}]{weigend05}%
  \BibitemOpen
  \bibfield  {author} {\bibinfo {author} {\bibfnamefont {F.}~\bibnamefont
  {Weigend}}\ and\ \bibinfo {author} {\bibfnamefont {R.}~\bibnamefont
  {Ahlrichs}},\ }\bibfield  {title} {\bibinfo {title} {Balanced basis sets of
  split valence, triple zeta valence and quadruple zeta valence quality for h
  to rn: Design and assessment of accuracy},\ }\href@noop {} {\bibfield
  {journal} {\bibinfo  {journal} {Physical Chemistry Chemical Physics}\
  }\textbf {\bibinfo {volume} {7}},\ \bibinfo {pages} {3297} (\bibinfo {year}
  {2005})}\BibitemShut {NoStop}%
\bibitem [{\citenamefont {Frisch}\ \emph {et~al.}(2016)\citenamefont {Frisch}
  \emph {et~al.}}]{frisch16}%
  \BibitemOpen
  \bibfield  {author} {\bibinfo {author} {\bibfnamefont {M.~J.}\ \bibnamefont
  {Frisch}} \emph {et~al.},\ }\href@noop {} {\bibinfo {title} {Gaussian~16
  {R}evision {C}.01}} (\bibinfo {year} {2016}),\ \bibinfo {note} {{G}aussian
  Inc. Wallingford CT}\BibitemShut {NoStop}%
\bibitem [{Note3()}]{Note3}%
  \BibitemOpen
  \bibinfo {note} {The $R$ distribution is proportional to the product of the
  degeneracy and the Boltzmann factors. The degeneracy of spherical top
  molecules like C$_{60}$ is broken due to ro-vibrational effects~\cite
  {bunker99}, giving a degeneracy of $2R+1$ (as for symmetric tops) rather than
  $(2R+1)^2$. The C$_{60}$ rotational constant is $\protect \tilde B$ =
  0.0028~cm$^{-1}$ ($\protect \tilde A$ = 0.0022~cm$^{-1}$ and $\protect \tilde
  B$ = 0.0019~cm$^{-1}$ for C$_{70}$).}\BibitemShut {Stop}%
\bibitem [{Note4()}]{Note4}%
  \BibitemOpen
  \bibinfo {note} {Averaging over the $R$ distribution at 870~K gives
  comparable results to using $R_{\protect \REV@text {max}}$, which is
  computationally more efficient.}\BibitemShut {Stop}%
\bibitem [{\citenamefont {Amirav}\ \emph {et~al.}(1980)\citenamefont {Amirav},
  \citenamefont {Even},\ and\ \citenamefont {Jortner}}]{amirav80}%
  \BibitemOpen
  \bibfield  {author} {\bibinfo {author} {\bibfnamefont {A.}~\bibnamefont
  {Amirav}}, \bibinfo {author} {\bibfnamefont {U.}~\bibnamefont {Even}},\ and\
  \bibinfo {author} {\bibfnamefont {J.}~\bibnamefont {Jortner}},\ }\bibfield
  {title} {\bibinfo {title} {Cooling of large and heavy molecules in seeded
  supersonic beams},\ }\href
  {https://doi.org/https://doi.org/10.1016/0301-0104(80)80077-2} {\bibfield
  {journal} {\bibinfo  {journal} {Chemical Physics}\ }\textbf {\bibinfo
  {volume} {51}},\ \bibinfo {pages} {31} (\bibinfo {year} {1980})}\BibitemShut
  {NoStop}%
\bibitem [{\citenamefont {Hornberger}\ \emph {et~al.}(2003)\citenamefont
  {Hornberger}, \citenamefont {Uttenthaler}, \citenamefont {Brezger},
  \citenamefont {Hackerm\"uller}, \citenamefont {Arndt},\ and\ \citenamefont
  {Zeilinger}}]{hornberger03}%
  \BibitemOpen
  \bibfield  {author} {\bibinfo {author} {\bibfnamefont {K.}~\bibnamefont
  {Hornberger}}, \bibinfo {author} {\bibfnamefont {S.}~\bibnamefont
  {Uttenthaler}}, \bibinfo {author} {\bibfnamefont {B.}~\bibnamefont
  {Brezger}}, \bibinfo {author} {\bibfnamefont {L.}~\bibnamefont
  {Hackerm\"uller}}, \bibinfo {author} {\bibfnamefont {M.}~\bibnamefont
  {Arndt}},\ and\ \bibinfo {author} {\bibfnamefont {A.}~\bibnamefont
  {Zeilinger}},\ }\bibfield  {title} {\bibinfo {title} {Collisional decoherence
  observed in matter wave interferometry},\ }\href
  {https://doi.org/10.1103/PhysRevLett.90.160401} {\bibfield  {journal}
  {\bibinfo  {journal} {Phys. Rev. Lett.}\ }\textbf {\bibinfo {volume} {90}},\
  \bibinfo {pages} {160401} (\bibinfo {year} {2003})}\BibitemShut {NoStop}%
\bibitem [{\citenamefont {Cheeseman}\ \emph {et~al.}(1996)\citenamefont
  {Cheeseman}, \citenamefont {Trucks}, \citenamefont {Keith},\ and\
  \citenamefont {Frisch}}]{cheeseman96}%
  \BibitemOpen
  \bibfield  {author} {\bibinfo {author} {\bibfnamefont {J.~R.}\ \bibnamefont
  {Cheeseman}}, \bibinfo {author} {\bibfnamefont {G.~W.}\ \bibnamefont
  {Trucks}}, \bibinfo {author} {\bibfnamefont {T.~A.}\ \bibnamefont {Keith}},\
  and\ \bibinfo {author} {\bibfnamefont {M.~J.}\ \bibnamefont {Frisch}},\
  }\bibfield  {title} {\bibinfo {title} {A comparison of models for calculating
  nuclear magnetic resonance shielding tensors},\ }\href
  {https://doi.org/10.1063/1.471789} {\bibfield  {journal} {\bibinfo  {journal}
  {The Journal of Chemical Physics}\ }\textbf {\bibinfo {volume} {104}},\
  \bibinfo {pages} {5497} (\bibinfo {year} {1996})}\BibitemShut {NoStop}%
\bibitem [{\citenamefont {Gauss}\ \emph {et~al.}(1996)\citenamefont {Gauss},
  \citenamefont {Ruud},\ and\ \citenamefont {Helgaker}}]{gauss96}%
  \BibitemOpen
  \bibfield  {author} {\bibinfo {author} {\bibfnamefont {J.}~\bibnamefont
  {Gauss}}, \bibinfo {author} {\bibfnamefont {K.}~\bibnamefont {Ruud}},\ and\
  \bibinfo {author} {\bibfnamefont {T.}~\bibnamefont {Helgaker}},\ }\bibfield
  {title} {\bibinfo {title} {Perturbation‐dependent atomic orbitals for the
  calculation of spin‐rotation constants and rotational g tensors},\ }\href
  {https://doi.org/10.1063/1.472143} {\bibfield  {journal} {\bibinfo  {journal}
  {The Journal of Chemical Physics}\ }\textbf {\bibinfo {volume} {105}},\
  \bibinfo {pages} {2804} (\bibinfo {year} {1996})}\BibitemShut {NoStop}%
\bibitem [{\citenamefont {Wilson}\ \emph {et~al.}(2005)\citenamefont {Wilson},
  \citenamefont {Mohn},\ and\ \citenamefont {Helgaker}}]{wilson05}%
  \BibitemOpen
  \bibfield  {author} {\bibinfo {author} {\bibfnamefont {D.~J.~D.}\
  \bibnamefont {Wilson}}, \bibinfo {author} {\bibfnamefont {C.~E.}\
  \bibnamefont {Mohn}},\ and\ \bibinfo {author} {\bibfnamefont
  {T.}~\bibnamefont {Helgaker}},\ }\bibfield  {title} {\bibinfo {title} {The
  rotational g tensor as a benchmark for density-functional theory calculations
  of molecular magnetic properties},\ }\href
  {https://doi.org/10.1021/ct050101t} {\bibfield  {journal} {\bibinfo
  {journal} {Journal of Chemical Theory and Computation}\ }\textbf {\bibinfo
  {volume} {1}},\ \bibinfo {pages} {877} (\bibinfo {year} {2005})}\BibitemShut
  {NoStop}%
\bibitem [{\citenamefont {Kendall}\ \emph {et~al.}(1992)\citenamefont
  {Kendall}, \citenamefont {Dunning},\ and\ \citenamefont
  {Harrison}}]{kendall92}%
  \BibitemOpen
  \bibfield  {author} {\bibinfo {author} {\bibfnamefont {R.~A.}\ \bibnamefont
  {Kendall}}, \bibinfo {author} {\bibfnamefont {T.~H.}\ \bibnamefont
  {Dunning}},\ and\ \bibinfo {author} {\bibfnamefont {R.~J.}\ \bibnamefont
  {Harrison}},\ }\bibfield  {title} {\bibinfo {title} {Electron affinities of
  the first‐row atoms revisited. systematic basis sets and wave functions},\
  }\href {https://doi.org/10.1063/1.462569} {\bibfield  {journal} {\bibinfo
  {journal} {The Journal of Chemical Physics}\ }\textbf {\bibinfo {volume}
  {96}},\ \bibinfo {pages} {6796} (\bibinfo {year} {1992})}\BibitemShut
  {NoStop}%
\bibitem [{\citenamefont {Storey}\ and\ \citenamefont
  {Cohen-Tannoudji}(1994)}]{Storey94}%
  \BibitemOpen
  \bibfield  {author} {\bibinfo {author} {\bibfnamefont {P.}~\bibnamefont
  {Storey}}\ and\ \bibinfo {author} {\bibfnamefont {C.}~\bibnamefont
  {Cohen-Tannoudji}},\ }\bibfield  {title} {\bibinfo {title} {The {F}eynman
  path integral approach to atomic interferometry. {A}~tutorial},\ }\href
  {https://doi.org/10.1051/jp2:1994103} {\bibfield  {journal} {\bibinfo
  {journal} {Journal de Physique {II}}\ }\textbf {\bibinfo {volume} {4}},\
  \bibinfo {pages} {1999} (\bibinfo {year} {1994})}\BibitemShut {NoStop}%
\bibitem [{\citenamefont {Bunker}\ and\ \citenamefont
  {Jensen}(1999)}]{bunker99}%
  \BibitemOpen
  \bibfield  {author} {\bibinfo {author} {\bibfnamefont {P.~R.}\ \bibnamefont
  {Bunker}}\ and\ \bibinfo {author} {\bibfnamefont {P.}~\bibnamefont
  {Jensen}},\ }\bibfield  {title} {\bibinfo {title} {Spherical top molecules
  and the molecular symmetry group},\ }\href
  {https://doi.org/10.1080/00268979909482827} {\bibfield  {journal} {\bibinfo
  {journal} {Molecular Physics}\ }\textbf {\bibinfo {volume} {97}},\ \bibinfo
  {pages} {255} (\bibinfo {year} {1999})}\BibitemShut {NoStop}%
\end{thebibliography}%

\end{document}


\title{Supplemental Material\\ Nanoscale magnetism probed in a matter-wave interferometer}

\author{Yaakov Y. Fein}
\email{yaakov.fein@univie.ac.at}
\affiliation{University of Vienna, Faculty of Physics, Vienna Center for Quantum Science and Technology (VCQ), Boltzmanngasse 5, A-1090 Vienna, Austria\\}
\author{Sebastian Pedalino}
\affiliation{University of Vienna, Faculty of Physics, Vienna Center for Quantum Science and Technology (VCQ), Boltzmanngasse 5, A-1090 Vienna, Austria\\}
\affiliation{University of Vienna, Vienna Doctoral School in Physics, Boltzmanngasse 5, A-1090 Vienna, Austria\\}
\author{Armin Shayeghi}
\affiliation{University of Vienna, Faculty of Physics, Vienna Center for Quantum Science and Technology (VCQ), Boltzmanngasse 5, A-1090 Vienna, Austria\\}
\author{Filip Kiałka}
\affiliation{University of Vienna, Faculty of Physics, Vienna Center for Quantum Science and Technology (VCQ), Boltzmanngasse 5, A-1090 Vienna, Austria\\}
\author{Stefan Gerlich}
\affiliation{University of Vienna, Faculty of Physics, Vienna Center for Quantum Science and Technology (VCQ), Boltzmanngasse 5, A-1090 Vienna, Austria\\}
\author{Markus Arndt}
\affiliation{University of Vienna, Faculty of Physics, Vienna Center for Quantum Science and Technology (VCQ), Boltzmanngasse 5, A-1090 Vienna, Austria\\}

\maketitle

\section{Experimental details}

\subsection{Anti-Helmholtz coils}
The anti-Helmholtz coils are composed of two circular coils each consisting of 52 turns of Kapton-insulated wire (2~mm diameter) wrapped in four layers around an aluminum frame. To obtain a homogenous gradient near the coil center we used a distance between the center of the coils of approximately $\sqrt 3 r = 0.07~m$, with $r=0.04$ the coil radius. The coil assembly was mounted in vacuum on an $x-y-z-\varphi$ manipulator to allow in-situ alignment of the pinholes with respect to the atomic beam. 
\begin{figure} 
    \includegraphics[width=\columnwidth]{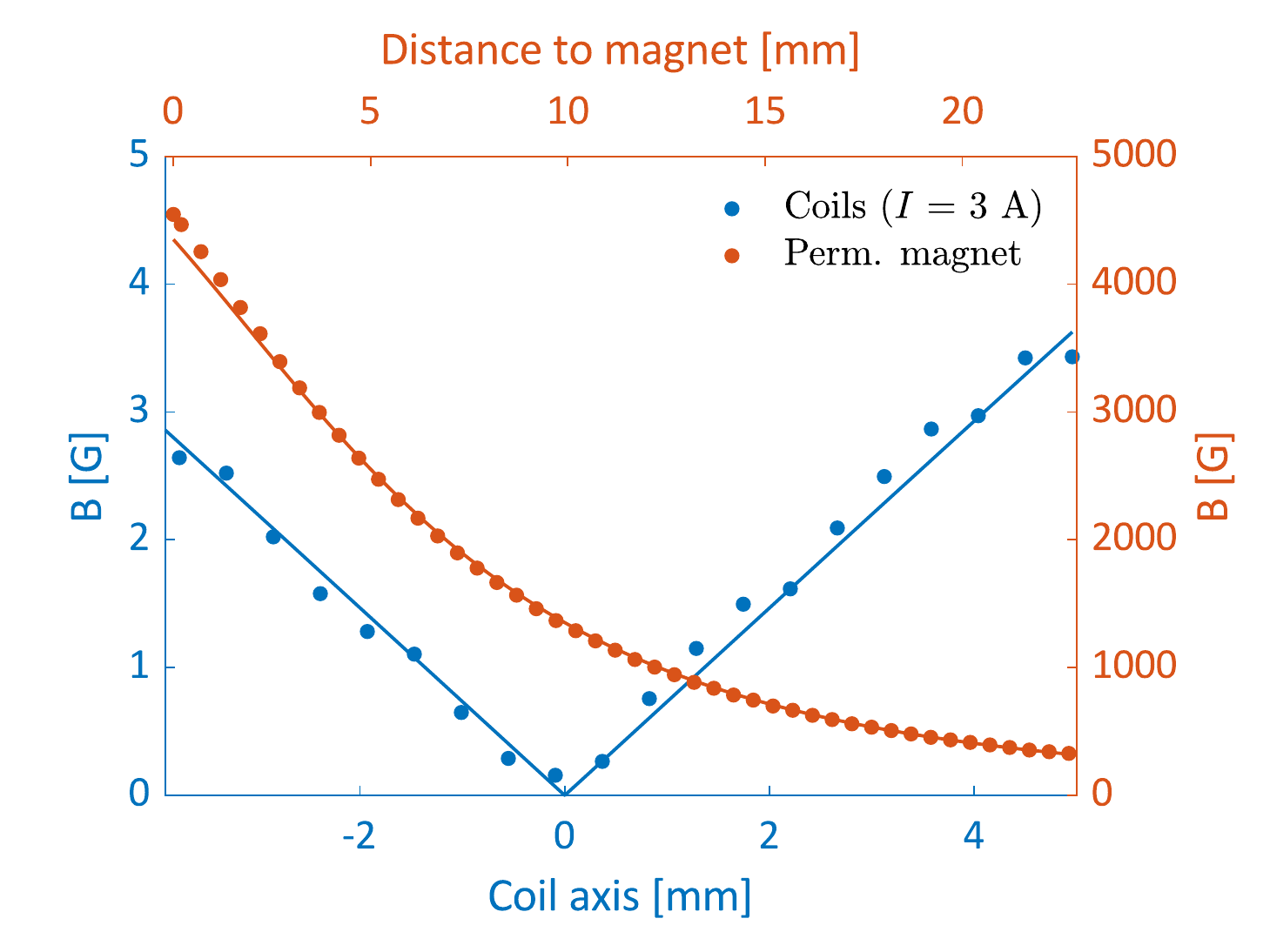}
    \caption{\label{fig:fields} Characterization of the magnetic fields produced by the anti-Helmholtz coils and permanent magnet. Left axis (blue): Magnetic fields of the coils at a current of 3~A along the coil axis ($y=z=0$). Right axis (red): On-axis $B$ of the permanent magnet as a function of distance to the surface. Solid lines are theory, as described in the text.}
\end{figure}
\begin{figure}
    \includegraphics[width=\columnwidth]{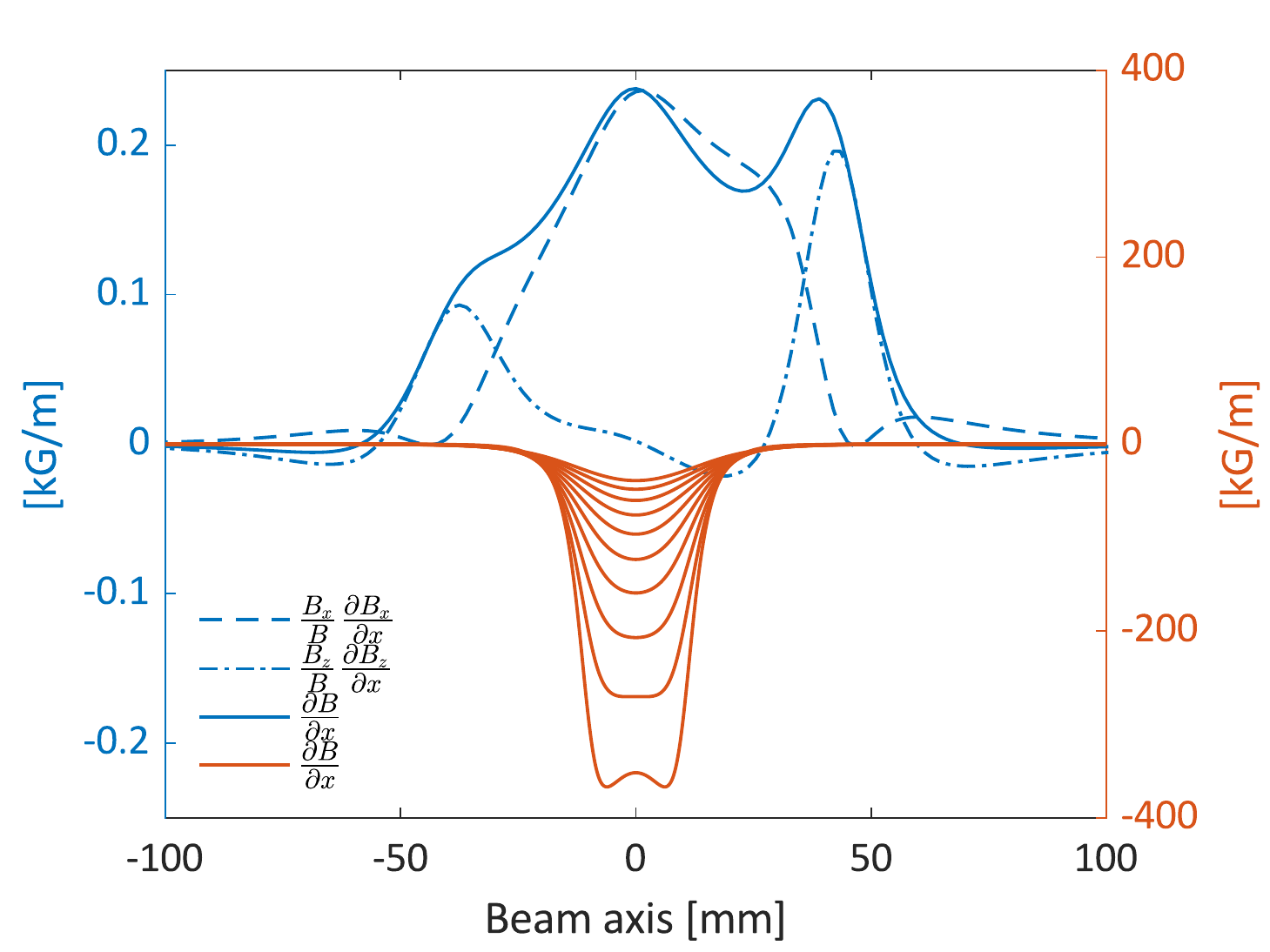}
    \caption{\label{fig:force} Left axis (blue): Simulation of the non-zero contributions to $\partial B/ \partial x$ for a current of 1~A using the magnetic field model evaluated along the atomic trajectories. The asymmetry of the coils gradient is due to the trajectory tilt. Right axis (red): $\partial B / \partial x$ for various distances from the permanent magnet surface at a vertical ($y$) offset of 6~mm (as during interference measurements). The maximal force (lowermost curve) is for a distance of 5~mm, minimal (uppermost curve) at a distance of 20~mm. Note the different scale of the right $y$-axis.}
\end{figure}

The fields produced by the coils were systematically mapped with a 3-axis Hall probe (Melexis MLX90393). A subset of this data is shown in Fig.~\ref{fig:fields}. We model the fields by numerically solving the Biot-Savart law for each of the 104 coil loops positioned according to their measured geometry~\cite{bergeman87}. There are no free parameters except for a 0.4~mm offset of the sensor head in the longitudinal direction as determined from the zero-field crossing point of $B_x$. Good theoretical agreement was also found for fields taken off the central axis, confirming the validity of the model.

The modeled field gradients, which enter via the $C$-factor in the expression for magnetic phase (Eqs.~2~and~3 of the main text), are shown in Fig.~\ref{fig:force}. In general, $\partial B /\partial x \neq \partial B_x /\partial x$, i.e., the various components of the gradient must be explicitly considered.

The atomic trajectories were defined by pinhole plates at the entry and exit of the coils. After the measurements, we determined that the trajectories were at a small angle (3.6°) with respect to the coils (i.e., not perfectly orthogonal to the coil axis), which is included in the analysis. For the coils, we use $L_1=0.30$~m and $L_2=0.24$~m to ensure that the force drops to a negligible value in the calculation of $C$.

\subsection{Permanent magnet}
A 2~cm~$\times$~2~cm~$\times$~1~cm NdFeB magnet was installed on a linear translation stage along the coil axis, keeping the lengths in Fig.~1 similar to those for the coil experiments. The magnet could be translated from a position grazing the molecular beam to a maximum of 8~cm away from the molecular beam, and has a specified 1.3~T remanent magnetization. This value was confirmed by measuring the on-axis $B$ with a F. W. Bell FH-520 Hall probe, as shown in Fig.~\ref{fig:fields}. As for the coils, the modeled gradients are shown in Fig.~\ref{fig:force}.

Since the form of the magnetic force depends on whether the magnetic moment is induced or permanent, we use two different lengths to ensure the force drops to a comparably small value in both cases in the calculation of $C$. For an induced moment ($F_x \propto (\bm B \cdot \nabla) B_x$), we use $L_1=0.06$~m and $L_2=0.37$~m, while for a permanent moment ($F_x \propto \partial B / \partial x$), we use $L_1=0.15$~m and $L_2=0.32$~m. 

\begin{figure}
    \includegraphics[width=\columnwidth]{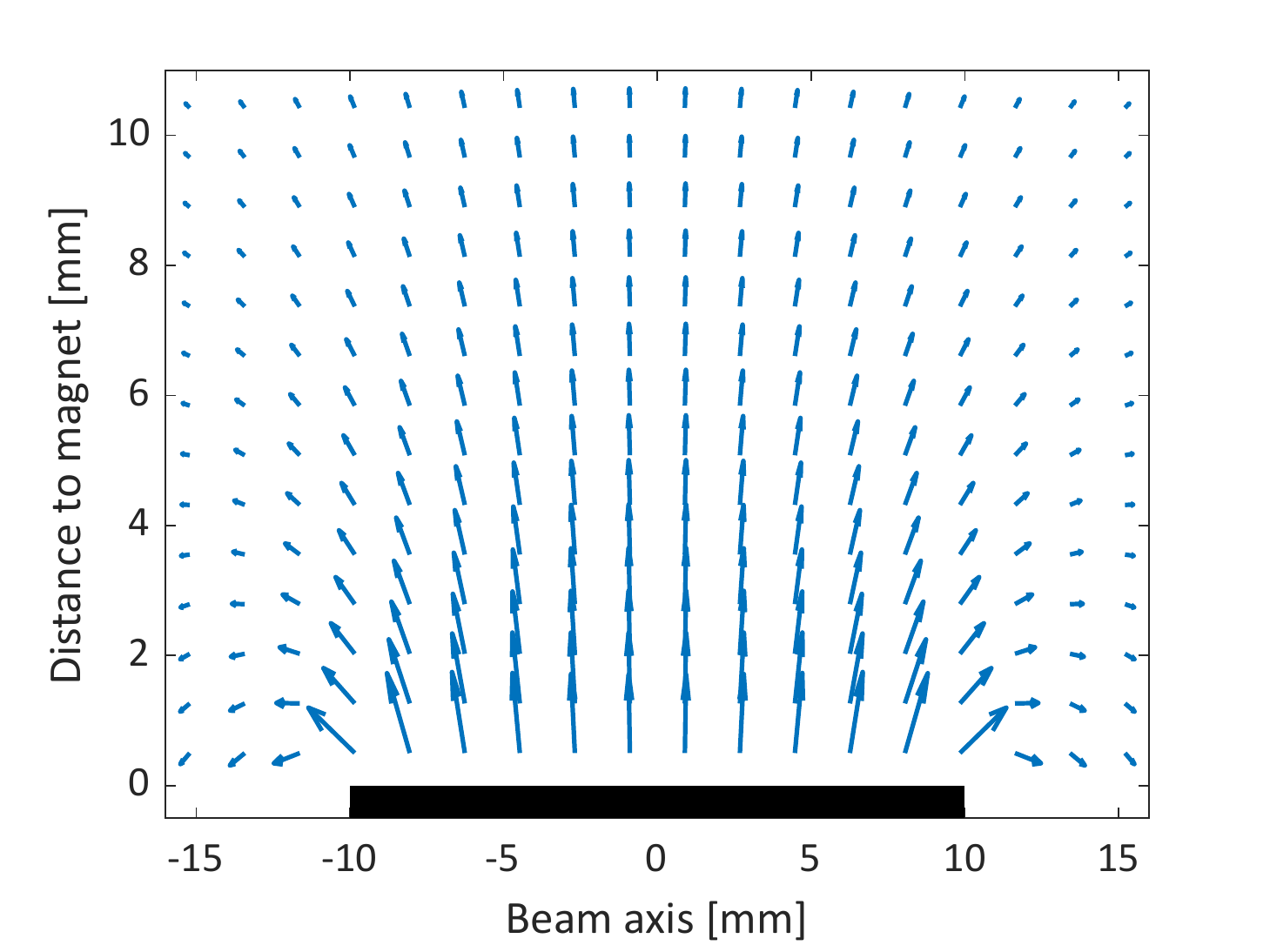}
    \caption{\label{fig:perm_rot} Simulated field components $B_x$ and $B_z$ above the permanent magnet.}
\end{figure}

The direction of $\bm B$ as the molecules traverse the permanent magnet is important for understanding the reduced effect of rotational magnetic moments on symmetric top molecules like C$_{70}$. The rotation of $\bm B$ can be clearly seen in Fig.~\ref{fig:perm_rot}. 

\subsection{Detection, velocity measurement, and fitting}
After the third grating, the atoms are surface ionized with a heated rhenium filament and the molecules are ionized via electron bombardment. The ions are then deflected into a quadrupole mass spectrometer which filters out background counts and can be used for isotope selection. Velocity measurements of the atoms and fullerenes are made by modulating the beam with a mechanical chopper in the beamline, while a time-of-flight distribution could be directly measured for the pulsed beam of radicals. 

Sinusoidal visibilities of the interference fringes were extracted by fitting each interference scan after subtraction of the detector dark rate. In the atomic experiments, a robust (bisquare) fit was used to reduce the influence of outliers caused by signal spikes from the hot wire detector. 

\subsection{Background fields}
A constant background field gradient $\partial B_0/\partial x$ over the length of the interferometer would give a contribution $C_0 = L^2 \partial B_0 /\partial x$ to the magnetic phase shift. The values of 0.4~G/m and 1.3~G/m for the cesium and rubidium measurements respectively were determined via least-squares fitting of the visibility response and are consistent with realistic background gradients. 

While $C_0$ describes the effect of a background gradient, constant background fields $B_0$ also play a role at low currents. When $B_0$ exceeds the field produced by the coils (which occurs at around 0.1~A for typical background fields of $\approx 0.4$~G), the magnetic moments are re-oriented along an axis not solely defined by the coil fields. This explains the discrepancy with theory at these very low current values. 

\subsection{Cesium}
For the cesium (Sigma-Aldrich, $\geq 99.95\%$) data labeled as 380~m/s, four interference scans with an integration time of one second per transverse position were conducted at each current setting (except at 0.9, 1.1, 1.3 and 1.5~A, for which two scans were collected). The QMS could be operated in a low-resolution mode to optimize flux since cesium has only one stable isotope and was thus mainly used to filter out the low-mass outgassing of metal ions from the heated rhenium wire used for ionization. 

At currents beyond 5~A, the visibility dropped below the asymptote, as seen in the inset of Fig.~2 in the main text. Since in this current regime we observed a temperature rise of the coil assembly and a corresponding pressure rise in the vacuum chamber, we speculate that the visibility drop could be due to collisional decoherence~\cite{hornberger03} caused by outgassing of the Kapton insulation of the coil wires. The observation of similar behavior for a range of species with different magnetic properties further suggests that the high-current regime is dominated by a non-magnetic systematic effect. 

The 270~m/s cesium data was achieved by gravitationally selecting the slower trajectories via vertical delimiters in the beamline. For this measurement, three interference scans with an integration time of one second per transverse position were conducted at each current setting and the QMS was operated at a higher resolution for signal stability reasons. Although we label the cesium measurements by their peak velocities in the main text, the velocity measurements show that the distributions are in fact significantly skewed. Fitting to a skew normal distribution as in Ref.~\cite{fein20} yields location parameters of 228~(290)~m/s, scale parameters of 118~(171)~m/s and shape parameters of 4.4 (2.1) for the 270~(380)~m/s datasets.

Since the zero-current points do not reach full visibility (13\% and 20\% for the slow and fast datasets respectively), we normalize the visibilities to the asymptotes rather than to the zero-current visibility to avoid introducing a scaling error. This is done by setting ${V_0=N \bar V(I_{\text{asympt}})/2}$, with $I_{\text{asympt}}$ taken as the range 4 to 4.5~A.

\subsection{Rubidium}
We performed analogous experiments with two isotopes of rubidium (Sigma-Aldrich, $\geq 99.6\%$) to confirm the qualitatively different response expected for atoms with a different hyperfine structure. We studied the isotopes $^{85}$Rb and $^{87}$Rb, for which we expect asymptote levels of 2/12 and 2/8 respectively (see Table~\ref{tab:1}). 
\begin{table}
    \caption{\label{tab:1} Hyperfine structure and magnetic moment projections ($\mu_z$, where $z$ is the quantization axis, not to be confused with the space-fixed $z$-axis of the main text) of the three isotopes.}
    \begin{ruledtabular}
    \begin{tabular}{ccccccccc}
     \multicolumn{3}{c}{$^{133}$Cs} &\multicolumn{3}{c}{$^{85}$Rb} &\multicolumn{3}{c}{$^{87}$Rb}\\
     $F$ &$m_F$ &$\mu_{z}$ 
     &$F$ &$m_F$ &$\mu_{z}$ 
     &$F$ &$m_F$ &$\mu_{z}$ \\ \hline
     3 &0 &0 &2 &0 &0 &1 &0 &0\\
     3 &$\pm1$ &$\mp \mu_B/4$ &2 &$\pm1$ &$\mp \mu_B/3$ &1 &$\pm1$ &$\mp \mu_B/2$\\
     3 &$\pm2$ &$\mp \mu_B/2$ &2 &$\pm2$ &$\mp 2\mu_B/3$ &2 &0 &0\\
     3 &$\pm3$ &$\mp 3\mu_B/4$ &3 &0 &0 &2 &$\pm1$ &$\pm \mu_B/2$\\
     4 &0 &0 &3 &$\pm1$ &$\pm \mu_B/3$ &2 &$\pm2$ &$\pm \mu_B$\\
     4 &$\pm1$ &$\pm \mu_B/4$ &3 &$\pm2$ &$\pm 2\mu_B/3$ &-- &-- &--\\
     4 &$\pm2$ &$\pm \mu_B/2$ &3 &$\pm3$ &$\pm \mu_B$ &-- &-- &--\\
     4 &$\pm3$ &$\pm 3\mu_B/4$ &-- &-- &-- &-- &-- &--\\
     4 &$\pm4$ &$\pm \mu_B$ &-- &-- &-- &-- &-- &--\\
    \end{tabular}
    \end{ruledtabular}
\end{table}

\begin{figure}
    \includegraphics[width=\columnwidth]{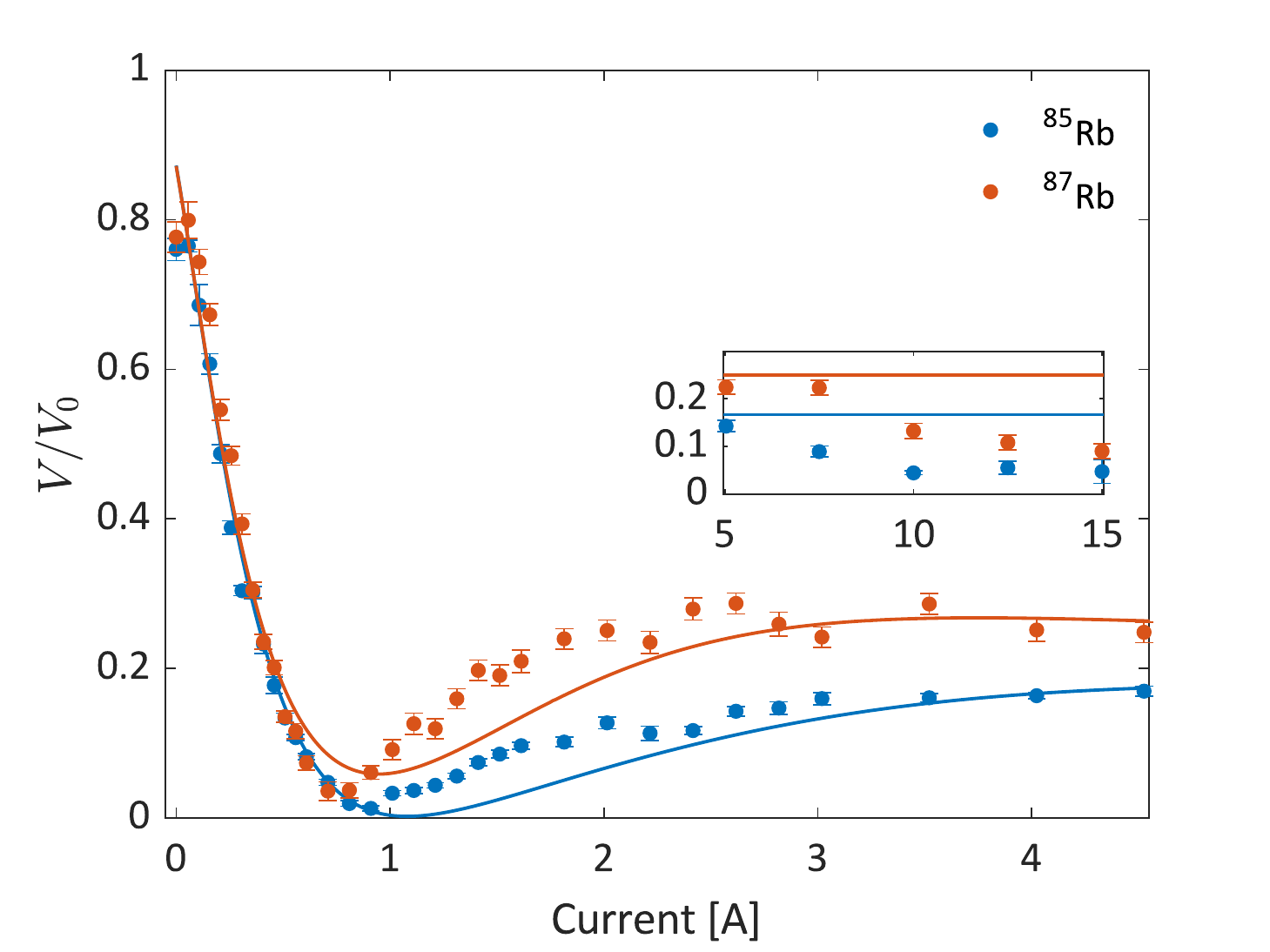}
    \caption{\label{fig:data_rb} Rubidium interference visibility as a function of anti-Helmholtz coil current. The $C$ factor in the solid theory curves is determined by the coil geometry, with an additional constant background gradient of 1.3~G/m. Each point consists of multiple interference scans, and the error bars are standard errors. The inset shows the high-current behavior, with the line indicating the expected asymptote.}
\end{figure}

For the rubidium measurements we operated the QMS in a high-resolution mode to differentiate the isotopes, which also limited the flux compared to the cesium measurements. Maximum visibility of 19\% was obtained for both isotopes, and the magnetic response is shown in Fig.~\ref{fig:data_rb}. For $^{85}$Rb, three interference scans with one second integration time were conducted per current setting until 5~A, and four scans at higher currents. Since $^{87}$Rb has a smaller isotopic abundance, more interference scans per current setting were required to obtain a similar statistical uncertainty and to compensate for flux instability encountered from about 1~A onwards. Five scans with one second integration time were made in the region 0 to 0.9~A and 7.5 to 15~A, while six were made from 1 to 5~A. 

We fit the velocity to a skew normal distribution, and find a location parameter of 425~m/s, a scale parameter of 220~m/s and a shape parameter of 1.7. The rubidium isotopes were measured in consecutive data collection runs and share the same velocity distribution, as measured at the beginning and end of the entire measurement series. However, we speculate that flux instabilities and a change in source temperature during the $^{87}$Rb series may correlate with higher-than-expected velocities in the region past 1~A, which would explain the theoretical disagreement in this region. A small geometric offset of the coils compared with the cesium measurements may also play a role in the poorer agreement with the theory as compared with the cesium data.

\subsection{TEMPO}
The TEMPO (Sigma-Aldrich, 99\%) beam was formed in a supersonic expansion of argon at a backing pressure of 2.5~bar (relative to vacuum). A pulse length of 28~µs at 100~Hz was used for the valve~\cite{even00}. The molecule cartridge was heated to 350~K while the valve mechanism was heated to 420~K. With 19.2~W in the phase grating, we could achieve a baseline visibility of 9\%, limited by TEMPO's low optical polarizability at 532 nm and its high forward velocity. 

\subsection{Fullerene isotopomers}
\begin{figure}
    \includegraphics[width=\columnwidth]{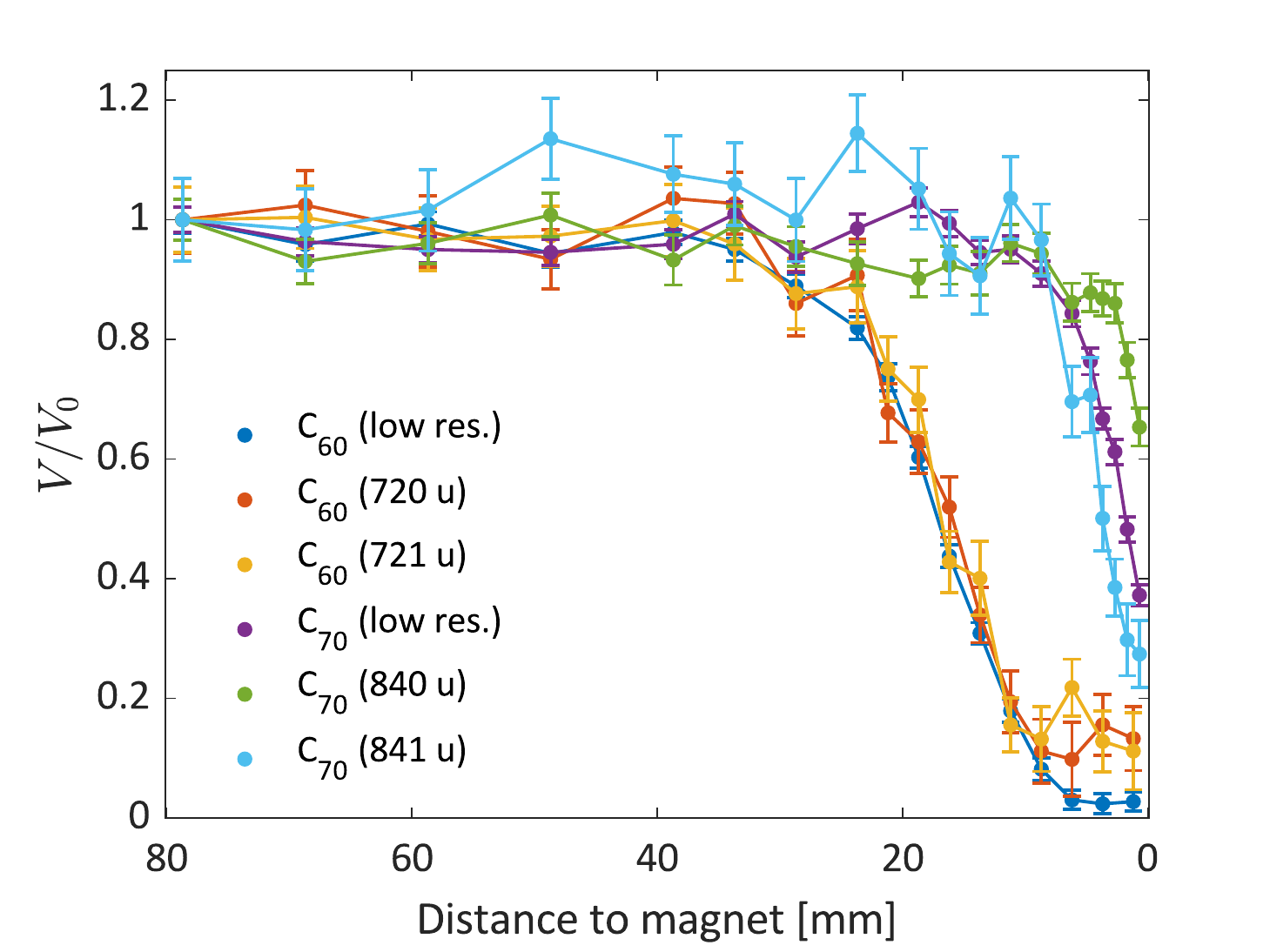}
    \caption{\label{fig:data_isotopes} Fullerene interference visibility as a function of magnet distance, for various isotopomers. Visibilities are normalized to the first data point (magnet withdrawn) and error bars are standard errors. Connecting lines are drawn to guide the eye.}
\end{figure}

We studied the magnetic response of the fullerene isotopomers (MER, $\geq 99\%$ purity) independently, as shown in Fig.~\ref{fig:data_isotopes}. $^{12}$C$_{69}$$^{13}$C  exhibits a stronger magnetic response, which affects the low-resolution C$_{70}$ data due to the large proportion of $^{13}$C-containing molecules in an isotopically-unresolved sample of C$_{70}$. The C$_{60}$ response, on the other hand, is identical within the error bars for all isotopomers. 

The maximum visibility for the C$_{60}$ measurements was 28\%, while the effects of photon absorption limited the  $^{12}$C$_{70}$ visibility to 16\% for $^{12}$C$_{70}$.

\section{Rotational magnetic moments}
\subsection{Symmetric top molecules}
The projection of a symmetric top's rotational magnetic moment onto a space-fixed axis is given by~\cite{eshbach52}
\begin{equation}
    \label{eq:muRot}
    \mu_{\text{rot}} = M \Big[g_{xx} + (g_{zz}-g_{xx})\frac{K^2}{R(R+1)} \Big] \mu_N.
\end{equation}
The $g_{\text{rot}}$ as referenced in the main text refers to the bracketed part of Eq.~\ref{eq:muRot} and hence depends in general on $K$ and $R$. For a spherical top this dependence is eliminated, since $g_{xx} = g_{zz}$, leading to $\mu_{\text{rot}} = M g_{xx} \mu_N$.

\subsection{Computational details}
The rotational $g$-tensor arises from the interaction of rotationally induced magnetic moments with external magnetic fields in what is sometimes referred to as the rotational Zeeman effect~\cite{flygare74}. Computationally, it is convenient to express the $g$-tensor as the second derivative of the molecular electronic energy with respect to the magnetic field strength and the rotational angular momentum. This way, $g_{\text{rot}}$ can be calculated like other time-independent second-order properties, when the expression for the energy in the presence of an external magnetic field is known. 

The calculation of $g_{\text{rot}}$, however, suffers from the gauge-origin problem and slow basis-set convergence. A standard approach for alleviating the gauge-origin problem is known as the gauge-including atomic orbital (GIAO) or London atomic orbital (LAO) approach~\cite{cheeseman96}. Gauss et al.~\cite{gauss96} introduced an efficient scheme for the calculation of $g_{\text{rot}}$ which ensures gauge-origin independent results and accelerates the convergence toward the basis-set limit.

Density functional theory for the calculation of $g_{\text{rot}}$ has been benchmarked~\cite{wilson05}, showing an overall good agreement with experimental data for a variety of molecules using the Becke three-parameter Lee-Yang-Parr (B3LYP) functional~\cite{becke93}. Since the use of diffuse functions has been shown to be advantageous~\cite{wilson05,gauss96}, we compared calculations for C$_{60}$ using Dunning’s correlation-consistent basis sets aug-cc-pVDZ and aug-cc-pVTZ~\cite{kendall92} with the Karlsruhe basis sets def2-TZVPP, def2-QZVPP and def2-TZVPPD~\cite{weigend05}. While the Dunning basis sets yield slow convergence, off-diagonal elements and non-identical diagonal elements, the Karlsruhe basis sets show good convergence and also agree better with the experiment.

The $g_{\text{rot}}$ values presented in this work were therefore calculated at the B3LYP/def2-TZVPP level of theory employing GIAOs within the Gaussian 16 program package~\cite{frisch16} as the best tradeoff between cost and accuracy. The geometries of TEMPO, C$_{60}$ and C$_{70}$ used in all calculations were optimized at the B3LYP/def2-TZVPP level of theory. The resulting $g_{\text{rot}}$ tensors are
\begin{align*}
    g_{\text{rot}}(\text{TEMPO}) &= \begin{pmatrix*}[r]
         -0.0098 & \hphantom{-0000.}0 & \hphantom{-0000.}0 \\
         0 & -0.0039 & 0 \\
         0 & 0 & -0.0104
         \end{pmatrix*}, \\
    g_{\text{rot}}(\text{C}_{60}) &= \begin{pmatrix*}[r] 
        -0.0141 & \hphantom{-0000.}0 & \hphantom{-0000.}0 \\
         0 & -0.0141 & 0 \\
         0 & 0 & -0.0141
         \end{pmatrix*}, \\
    g_{\text{rot}}(\text{C}_{70}) &= \begin{pmatrix*}[r] 
         \hphantom{-}0.0025 & \hphantom{-0000.}0 & \hphantom{-0000.}0 \\
          0 & 0.0025 & 0 \\
          0 & 0 & -0.0046
          \end{pmatrix*}. \\
\end{align*}

We have additionally calculated $g_{\text{rot}}$ for C$_{60}$ at distorted geometries taken from displacement patterns based on the $T_{\text{1u}}$ modes of C$_{60}$ (which are excited at the experimental temperatures) and confirmed that the  $g_{\text{rot}}$ values vary by less than 3\% when compared to the ground state geometry, justifying the use of the latter. We have also neglected zero-point vibrational corrections.

%